\documentclass{aastex631}
\usepackage{csquotes}
\begin{document}
\title{Radial Stellar Age Gradients in 42 Local Volume Dwarf Galaxies}

\correspondingauthor{Roger E. Cohen}
\email{rc1273@physics.rutgers.edu}

\author[0000-0002-2970-7435]{Roger E. Cohen}
\affiliation{Department of Physics and Astronomy, Rutgers the State University of New Jersey, 136 Frelinghuysen Rd., Piscataway, NJ, 08854, USA}
\email{rc1273@physics.rutgers.edu}
\affiliation{Eureka Scientific Inc., 2452 Delmer Street, Oakland, CA 94602, USA}

\author[0000-0001-5538-2614]{Kristen B. W. McQuinn}
\affiliation{Department of Physics and Astronomy, Rutgers the State University of New Jersey, 136 Frelinghuysen Rd., Piscataway, NJ, 08854, USA}
\affiliation{Space Telescope Science Institute, 3700 San Martin Drive, Baltimore, MD 21218, USA}

\author[0000-0003-4122-7749]{O. Grace Telford}
\affiliation{Department of Physics and Astronomy, University of Utah, 270 S 1400 E, Salt Lake City, UT 84112, USA}

\author{Liese van Zee}
\affiliation{Department of Astronomy, Indiana University, 727 East 3rd Street, Bloomington, IN 47405, USA}

\author[0000-0001-5368-3632]{Laura Congreve Hunter}
\affiliation{Department of Physics and Astronomy, Dartmouth College, 17 Fayerweather Hill Rd., Hanover, NH 03755, USA}

\author[0000-0001-8416-4093]{Andrew E. Dolphin}
\affiliation{Raytheon, 1151 E. Hermans Rd.,
Tucson, AZ 85756}
\affiliation{Steward Observatory, University of Arizona, 933 N. Cherry Avenue, Tucson, AZ 85719, USA}

\author[0009-0007-6658-0318]{Suchindram Dasgupta}
\affiliation{Department of Physics and Astronomy, West Virginia University, Morgantown, WV 26506, USA}

\begin{abstract}
    
We present radial stellar age gradients measured from star formation histories (SFHs) fit to resolved color-magnitude diagrams (CMDs) of 42 Local Volume dwarfs (6$\lesssim$Log M$_{\star}$/M$_{\odot}$$\lesssim$9), spatially divided into elliptical annuli.  Ages in each annulus are quantified using $\tau_{90}$ and  $\tau_{50}$, the lookback times to form 90\% and 50\% of the cumulative stellar mass.  We find that radial age gradients are uncorrelated with environment, but gradients in $\tau_{90}$ are significantly correlated ($p$-values$\lesssim$0.001) with lifetime galaxy-wide (\enquote{global}) SFHs, in agreement with two independent cosmological zoom-in simulations.  For radial gradients of $\tau_{50}$, simulation predictions differ.  We demonstrate that given our large (N=42) and diverse observational sample, the strength of an observed correlation with global SFH is an actionable parameter to discriminate between different simulations with differing stellar feedback prescriptions.  Overall, our results support predictions that dwarfs form inside-out like their more massive counterparts, with a combination of feedback-driven outward radial stellar migration and increasing birth radii for young stellar populations yielding present-day stellar age gradients ranging from outside-in to flat.  In addition, the lack of a correlation between $\tau_{50}$ gradients and global SFH argues against recent star formation in radially outflowing gas.  We also discuss the impact of differences between our observational sample and samples available in the latest simulations, highlighting areas for future investigation.  

\end{abstract}

\section{Introduction \label{introsect}}

Dwarf galaxies, by definition, have relatively low stellar masses (6$\lesssim$Log M$_{\star}$/M$_{\odot}$$\lesssim$9) and correspondingly shallow potential wells compared to their more massive counterparts.  They are therefore particularly sensitive probes of both the internal and external phenomena driving galaxy evolution.  A powerful avenue for probing the evolution of dwarfs is provided by their star formation histories (SFHs), chronicling their mass assembly over cosmic time.  Within the Local Volume, the Hubble Space Telescope (HST) has provided SFHs by resolving individual stars for large samples of dwarfs \citep[e.g.,][]{dalcanton09,weisz11,weisz13,weisz14,mcquinn09,mcquinn10,mcquinn10b,starbirds,gallart15,savino23,savino25}, complementing ground-based imaging for the nearest targets \citep[e.g.,][]{rubele18,ruizlara20,mazzi21,massana22}.  SFHs fit to resolved color-magnitude diagrams (CMDs) have yielded a plethora of insights into the connection between the assembly history of Local Volume dwarfs and other global properties, including mass, morphology and environment.  Examples include the impact of environment on the quenching of star formation \citep{weisz15,savino25} as well as the position-morphology and morphology-density relations \citep{einasto74,vandenbergh94a,vandenbergh94b,weisz11,weisz11b}.

\subsection{Radial Stellar Age Gradients in Dwarfs: Observations}

Beyond the \textit{global} SFHs of dwarf galaxies, observations have provided an intriguing picture of SFH trends \textit{within} dwarf galaxies.  On one hand, most dwarfs studied so far appear to have \enquote{outside-in} age gradients, where recent or ongoing star formation is centrally concentrated, resulting in preferentially older ages for the stellar populations located in galaxy outskirts \citep{sarajedini97,minniti97,dohmpalmer98,gallagher98,lee99,sl99,aparicio00,aparicio00b,drozdovsky00,annibali03,hidalgo03,cannon03,alonsogarcia06,bernard07,dejong08,hidalgo08,mcquinn12,hidalgo13,delpino13,santana16,mcquinn17,bettinelli19,albers19,savino19,sacchi21,cohen25}.  However, exceptions do exist, including the dwarf irregular (dIrr) NGC 6822, which has a flat age gradient \citep{cannon12,fusco14} and, while somewhat more massive, M33 and the LMC, which have inverted \enquote{V-shaped} radial stellar age profiles \citep{williams09,cohen24a}.  

Yet, with few exceptions, our picture of radial stellar age trends within dwarf galaxies remains almost entirely \textit{qualitative} rather than \textit{quantitative}.  There have been only a handful of studies quantifying radial stellar age gradients in dwarf galaxies, generally targeting individual cases or small samples.  Specifically, in addition to the individual cases of flat or inverted gradients mentioned above, outside-in age gradients were measured from Hubble imaging of a sample of four relatively isolated Local Group dwarfs \citep{hidalgo13}, the SMC \citep{cohen24b}, and the blue compact dwarf (BCD) UGC 4483 \citep{sacchi21}, in addition to the study of WLM using a combination of Hubble and JWST imaging \citep{cohen25}.  Therefore, despite the broad qualitative finding that observed age gradients in dwarfs range from outside-in to flat, existing quantitative information on radial gradient slopes is both scarce and, with the exceptions of UGC 4483 and the SMC, biased towards low-mass dwarfs near the edge of the Local Group.  With such a small number of existing age gradient measurements sampling such a restricted environment, it is impossible to know the extent to which the observed range of age gradients is driven by internal versus external factors, or some combination of the two.  For example, simulations predict that massive hosts impact the \textit{global} star formation histories of dwarfs out to $>$3 Mpc \citep{christensen24}, but without measurements of age gradients in galaxies inhabiting more isolated environments, it is unclear whether environment has any impact on \textit{internal} gradients of stellar age within dwarfs.  More generally, we lack the observational constraints to assess whether the relative impact of external (i.e., environmental) versus internal drivers of age gradients in dwarfs is related to global galaxy properties such as mass, metallicity or global mass assembly history.

Here, we address the lack of both size and diversity among the sample of dwarfs with measured stellar age gradients, providing self-consistent radial stellar age gradient measurements for 42 dwarfs in the Local Volume.  The size and diversity of our sample (see Sect.~\ref{targetsect}) provides, for the first time, the opportunity to understand the drivers of age gradients in dwarfs from both observational and theoretical standpoints.  Observationally, intra-sample comparisons enable tests for statistically significant correlations between age gradient slopes and global galaxy properties (e.g., mass, metallicity, and global mass assembly history).  Importantly, by probing beyond the Local Group out into the Local Volume, we may also directly test for observed correlations between age gradients in dwarfs versus their environment.  The resultant correlations (or lack thereof; see Sect.~\ref{obscorrsect}) then provide critical clues regarding the drivers of the age gradients.  

\subsection{Radial Age Gradients in Dwarfs: Simulations \label{introsimsect}}

Our compilation of homogeneous age gradient measurements in dwarfs also allows for comparisons against multiple independent sets of zoom-in cosmological hydrodynamical simulations.  The stellar mass resolution attained by the latest simulations ($<$10$^{3}$ M$_{\odot}$ in some cases, e.g.,~\citealt{christensen24}) facilitates predictions for radial age gradient slopes in dwarfs.  However, aside from the individual case of WLM  \citep{cohen25}, these predictions remain entirely untested until now.  Fortunately, there are now at least two different sets of cosmological simulations that make testable predictions for radial age gradients as a function of global galaxy properties.  
Age gradient slopes for 26 isolated dwarfs with present-day stellar masses of 5.67$\leq$Log M$_{\star}$/M$_{\odot}$$\leq$8.73 from the FIRE-2 simulations \citep{hopkins18} were analyzed by \citet{graus19}.  Independently, \citet{riggs24} analyzed a larger sample of 72 dwarfs covering a broader range of stellar masses (5.26$\leq$Log M$_{\star}$/M$_{\odot}$$\leq$9.13).  Unlike \citet{graus19}, they were able to examine dwarfs over a range of environments by combining field dwarfs from the \enquote{Marvel-ous dwarfs} simulations \citep{munshi21,christensen24} with dwarfs in group environments from the \enquote{Near Mint DC Justice League} simulations \citep[][also see \citealt{applebaum21}]{bellovary19,akins21}.  

The \citet{graus19} and \citet{riggs24} simulations both predict that present-day age gradients in dwarfs range from outside-in to flat.  They further predict that these stellar age gradients are a consequence of stellar feedback-induced fluctuations in the shallow potential wells of dwarfs, preferentially driving older stars to more external radii over time.  These same potential well fluctuations affect dark matter as well as baryons, flattening the central dark matter density profile from cuspy to cored \citep[e.g.,][]{navarro96,governato12,brook15,burger22}.  Both simulations also concur that in globally younger dwarfs (i.e., those forming a larger fraction of their cumulative stellar mass at more recent times), recent star formation can occur at galactocentric radii that increase with time, counteracting long-term outward radial migration of older stars and flattening radial gradients.  However, the two sets of simulations differ in their stellar feedback implementations.  Unlike the \citet{riggs24} simulations, the \citet{graus19} simulations predict that the potential well fluctuations in dwarfs drive radial outflows (and later,  inflows) of star forming gas, imparting an initial radial velocity to stars formed there.  Because of the radial velocities imparted to stars at birth, these \enquote{breathing modes} impact the timescales of radial migration \citep{elbadry16}, resulting in a marked difference in predicted age gradient slopes at lookback times prior to $\sim$4 Gyr ago.  We demonstrate that this difference in predicted age gradient slopes is observable (see Sect.~\ref{t50simsect}), introducing radial stellar age gradients as a novel avenue to discriminate between different sets of simulations with different feedback implementations.  

This study is organized as follows: In Sect.~\ref{datasect}, we describe our dwarf galaxy sample, observations and photometry.  In Sect.~\ref{methodsect}, we present our methodology for measuring stellar age gradients from SFHs fit independently to spatially selected subsamples over a range of galactocentric radii in each target galaxy.  In Sect.~\ref{resultsect} we provide our measured age gradients and compare them with both existing measurements and predictions from multiple simulations.  Lastly, in Sect.~\ref{futuresect}, we summarize our findings and discuss avenues for further investigation.  Additional details are provided in Appendices~\ref{testsect}-\ref{indslopesect}, where we test the sensitivity of our measured age gradients to various assumptions in our fitting methodology (Appendix~\ref{testsect}), quantify correlations between our age gradient slopes and all of the galaxy-wide properties examined (Appendix~\ref{coefftabsect}), and illustrate our radial age gradient fits for all of our target galaxies (Appendix~\ref{indslopesect}).  

\section{Data \label{datasect}}

\subsection{Target Sample \label{targetsect}}

Our sample was culled from archival HST imaging of Local Volume gas-rich  dwarf galaxies (6$\lesssim$Log$_{\rm 10}$ (M$_{\star}$/M$_{\odot}$)$\lesssim$9).  From a practical standpoint, selection of target galaxies in which radial age gradients can be measured 
self-consistently 
from SFH fits to resolved stars is driven by a compromise between the need for SFH precision and the need for spatial coverage.  Specifically,
we required targets to have 
imaging in two broadband filters extending $\gtrsim$2 mag faintward of the tip of the red giant branch (TRGB) to usefully constrain the SFHs of our target galaxies at lookback times beyond several Gyr \citep[e.g.,][]{mcquinn10,starbirds}.  
At the same time, we also required imaging with spatial coverage extending to $>$4.4 times the disk scalelength $h_{r}$ (measured from ellipse fits to 3.6$\micron$ surface photometry; see Sect.~\ref{structuralsect}) to provide a sufficient radial baseline within each galaxy for measurement of radial gradients self-consistently across our sample.  

\begin{figure}
  \begin{minipage}[c]{0.5\textwidth}
    \includegraphics[width=\textwidth]{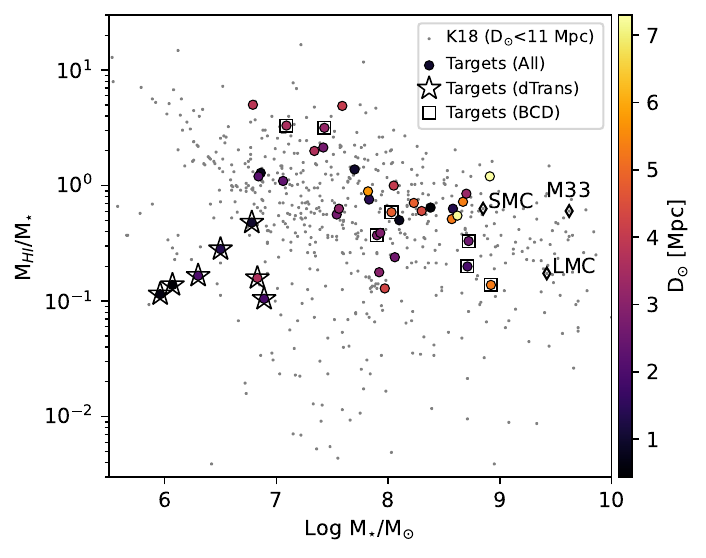}
  \end{minipage}\hfill
  \begin{minipage}[c]{0.45\textwidth}
    \caption{H\textsc{I}-to-stellar mass ratio versus stellar mass for Local Volume galaxies with D$_{\odot}$$\leq$11 Mpc from the \citet{k18} catalog, with our target galaxies overplotted and color-coded by distance.   Targets that are transition dwarfs (dTrans) or blue compact dwarfs (BCDs) rather than dwarf irregular (dIrr) galaxies are overplotted with star or square symbols respectively.  The loci of the LMC, SMC and M33 are indicated for comparison using diamonds and labeled.  Our targets span a broad ($>$3 dex) range in stellar mass and tend to be relatively gas-rich. \label{targetprops_fig}}
\end{minipage}
\end{figure}

The resulting sample consists of 42 Local Volume dwarfs with stellar masses 5.96$\leq$Log M$_{\star}$/M$_{\odot}$$\leq$8.92 and heliocentric distances 0.4$\lesssim$D$_{\odot}$$\lesssim$7.2 Mpc.  The vast majority of our targets lie at moderate to high Galactic latitudes (40/42 have $|B|$$>$20$^{\circ}$) and correspondingly low foreground extinction (40/42 have A$_{V}$$\leq$0.22 mag; \citealt{sf11}).   In Fig.~\ref{targetprops_fig} we plot the H\textsc{I}-to-stellar mass ratios of our target galaxies versus their stellar mass, illustrating that overall, our sample consists mainly of gas-rich dwarf irregular (dIrr) galaxies, although we have also included transition dwarfs (dTrans) 
to extend the low end of the stellar mass range we sample by an order of magnitude down to Log$_{\rm 10}$ M$_{\star}$/M$_{\odot}$$\sim$6.  The properties of our target galaxies are summarized in Table \ref{propstab}.

\begin{deluxetable}{llclccccccr}
\tabletypesize{\scriptsize}
\tablecaption{Target Galaxies: Global Properties \label{propstab}}
\tablehead{
\colhead{Galaxy} & \colhead{$\mathcal{A}$} & \colhead{A$_{V}$} & \colhead{D$_{\odot}$} & \colhead{Log $\frac{M_{\star}}{M_{\odot}}$} & \colhead{Log $\frac{M_{HI}}{M_{\odot}}$} & \colhead{12+Log(O/H)} & \colhead{$\theta_{1}$} & \colhead{D(NN)} & \colhead{D(NLG$_{10}$)} & Type \\ \colhead{} & \colhead{} & \colhead{mag} & \colhead{Mpc} & \colhead{} & \colhead{} & \colhead{dex} & \colhead{} & \colhead{Mpc} & \colhead{Mpc} & \colhead{}}
\startdata
WLM & 0.09 & 0.10 & 0.98$^{+0.02}_{-0.04}$\tablenotemark{a} & 7.70 & 7.84 & 7.77$\pm$0.10 & -0.04 & 0.22 & 0.86 & dI \\
ESO410-005 & 1.00 & 0.04 & 1.93$^{+0.03}_{-0.04}$ & 6.89 & 5.91 &  & 0.07 & 0.31 & 1.83 & dTr \\
ESO294-010 & 1.00 & 0.02 & 2.03$^{+0.03}_{-0.04}$ & 6.30 & 5.52 &  & 1.03 & 0.15 & 1.86 & dTr \\
ESO540-032 & 1.00 & 0.06 & 3.63$^{+0.05}_{-0.05}$ & 6.83 & 6.03 &  & 1.39 & 0.12 & 0.35 & dTr \\
LGS3 & 0.15 & 0.11 & 0.65$^{+0.14}_{-0.10}$ & 5.96 & 5.02 &  & 1.46 & 0.13 & 0.27 & dTr \\
IC1613 & 0.05 & 0.07 & 0.76$^{+0.02}_{-0.01}$ & 8.10 & 7.80 & 7.62$\pm$0.05 & 0.63 & 0.22 & 0.52 & dI \\
UGC685 & 0.89 & 0.16 & 4.81$^{+0.04}_{-0.04}$ & 8.03 & 7.80 & 8.00$\pm$0.03 & -1.51 & 1.48 & 3.24 & BCD \\
UGC1281 & 0.52 & 0.13 & 5.27$^{+0.05}_{-0.02}$ & 8.57 & 8.28 & 7.78$\pm$0.10 & -1.20 & 0.43 & 3.43 & dI \\
Phoenix & 0.19 & 0.04 & 0.42$^{+0.02}_{-0.02}$ & 6.07 & 5.21 &  & 0.67 & 0.16 & 0.44 & dTr \\
NGC0784 & 0.55 & 0.16 & 5.37$^{+0.05}_{-0.02}$ & 8.67 & 8.53 & 7.97$\pm$0.05 & -1.30 & 0.29 & 3.68 & dI \\
NGC2366 & 0.41 & 0.10 & 3.28$^{+0.05}_{-0.03}$ & 8.70 & 8.63 & 7.91$\pm$0.05 & 0.83 & 0.15 & 0.83 & dI \\
UGC4459 & 0.90 & 0.10 & 3.68$^{+0.05}_{-0.03}$ & 7.34 & 7.64 & 7.82$\pm$0.09 & 0.81 & 0.25 & 0.53 & dI \\
UGC4483 & 1.00 & 0.09 & 3.58$^{+0.15}_{-0.15}$ & 7.09 & 7.61 & 7.56$\pm$0.03 & 1.01 & 0.16 & 0.43 & dI/BCD \\
UGC5139 & 0.39 & 0.14 & 4.02$^{+0.04}_{-0.06}$ & 8.05 & 8.05 & 7.92$\pm$0.05 & 1.31 & 0.19 & 0.36 & dI \\
UGC5364 & 0.46 & 0.06 & 0.74$^{+0.16}_{-0.05}$ & 6.86 & 6.97 & 7.30$\pm$0.05 & -0.03 & 0.35 & 0.74 & dI \\
SextansB & 0.09 & 0.09 & 1.43$^{+0.02}_{-0.02}$ & 7.83 & 7.71 & 7.84$\pm$0.05 & -0.88 & 0.26 & 1.43 & dI \\
NGC3109 & 0.20 & 0.18 & 1.34$^{+0.06}_{-0.05}$ & 8.58 & 8.38 & 7.77$\pm$0.07 & -0.33 & 0.04 & 1.34 & dI \\
Antlia & 0.39 & 0.22 & 1.34$^{+0.08}_{-0.04}$ & 6.50 & 5.95 &  & 1.75 & 0.04 & 1.37 & dTr \\
SextansA & 0.12 & 0.12 & 1.45$^{+0.05}_{-0.05}$ & 7.57 & 7.90 & 7.54$\pm$0.09 & -0.89 & 0.26 & 1.45 & dI \\
NGC3738 & 0.86 & 0.03 & 5.30$^{+0.05}_{-0.05}$ & 8.92 & 8.06 & 8.04$\pm$0.06 & -1.01 & 0.36 & 1.78 & BCD \\
NGC3741 & 0.99 & 0.07 & 3.22$^{+0.16}_{-0.18}$ & 7.43 & 7.93 & 7.68$\pm$0.03 & -0.69 & 0.66 & 1.51 & BCD \\
UGC6817 & 0.24 & 0.07 & 2.65$^{+0.10}_{-0.10}$ & 7.42 & 7.75 & 7.53$\pm$0.02 & -0.62 & 0.30 & 1.89 & dI \\
NGC4068 & 0.85 & 0.06 & 4.39$^{+0.04}_{-0.04}$ & 8.30 & 8.08 & 8.00$\pm$0.20 & -0.48 & 0.25 & 1.07 & dI \\
NGC4163 & 1.00 & 0.06 & 2.99$^{+0.04}_{-0.03}$ & 7.92 & 7.17 & 7.56$\pm$0.14 & 1.08 & 0.03 & 1.53 & dI \\
NGC4190 & 0.28 & 0.08 & 2.83$^{+0.08}_{-0.08}$ & 7.90 & 7.47 & 7.93$\pm$0.20 & 1.72 & 0.06 & 1.66 & BCD \\
UGCA281 & 1.00 & 0.04 & 5.70$^{+0.11}_{-0.13}$ & 7.82 & 7.77 & 7.80$\pm$0.03 & -0.92 & 0.57 & 1.49 & dI \\
UGC7577 & 0.17 & 0.06 & 2.61$^{+0.05}_{-0.06}$ & 8.06 & 7.44 & 7.97$\pm$0.06 & -0.94 & 0.38 & 1.82 & dI \\
UGCA292 & 1.00 & 0.04 & 3.85$^{+0.55}_{-0.09}$ & 6.79 & 7.49 & 7.30$\pm$0.03 & -0.16 & 0.42 & 0.84 & dI \\
UGC8024 & 0.87 & 0.03 & 4.04$^{+0.07}_{-0.06}$ & 7.59 & 8.28 & 7.67$\pm$0.06 & 0.31 & 0.49 & 0.55 & dI \\
UGC8091 & 0.87 & 0.07 & 2.19$^{+0.09}_{-0.12}$ & 6.84 & 6.92 & 7.65$\pm$0.06 & -1.43 & 0.80 & 2.19 & dI \\
UGCA320 & 0.71 & 0.22 & 6.02$^{+0.28}_{-0.16}$ & 8.91 & 8.99 & 8.08$\pm$0.20 & 1.47 & 0.04 & 2.54 & dI \\
UGC8201 & 0.63 & 0.07 & 4.83$^{+0.04}_{-0.04}$ & 8.23 & 8.08 & 7.80$\pm$0.06 & -0.64 & 0.57 & 1.58 & dI \\
UGC8508 & 0.81 & 0.04 & 2.67$^{+0.10}_{-0.10}$ & 7.54 & 7.29 & 7.76$\pm$0.07 & -0.80 & 0.67 & 1.72 & dI \\
UGC8638 & 0.99 & 0.04 & 4.29$^{+0.06}_{-0.04}$ & 7.97 & 7.08 & 7.94$\pm$0.05 & -0.16 & 0.68 & 0.79 & dI \\
UGC8651 & 0.76 & 0.02 & 3.10$^{+0.06}_{-0.06}$ & 7.56 & 7.36 & 7.85$\pm$0.04 & -0.87 & 0.28 & 1.44 & dI \\
UGC8837 & 0.58 & 0.04 & 7.24$^{+0.03}_{-0.07}$ & 8.62 & 8.36 & 7.70$\pm$0.30 & 1.25 & 0.30 & 0.30 & dI \\
UGC9128 & 0.91 & 0.06 & 2.30$^{+0.03}_{-0.04}$ & 7.06 & 7.10 & 7.75$\pm$0.05 & -1.49 & 0.49 & 2.30 & dI \\
UGC9240 & 0.82 & 0.03 & 2.83$^{+0.05}_{-0.04}$ & 7.93 & 7.52 & 7.95$\pm$0.03 & -1.18 & 0.54 & 1.91 & dI \\
IC4662 & 1.00 & 0.19 & 2.55$^{+0.02}_{-0.02}$ & 8.72 & 8.24 & 8.17$\pm$0.04 & -1.24 & 1.12 & 1.20 & dI/BCD \\
NGC6822 & 0.06 & 0.56 & 0.51$^{+0.01}_{-0.01}$\tablenotemark{a} & 8.38 & 8.19 & 8.02$\pm$0.05 & 0.46 & 0.40 & 0.51 & dI \\
DDO210 & 0.34 & 0.14 & 0.98$^{+0.07}_{-0.06}$ & 6.78 & 6.46 &  & -0.35 & 0.36 & 0.97 & dTr \\
IC5152 & 0.35 & 0.07 & 1.96$^{+0.03}_{-0.05}$ & 8.71 & 8.01 & 7.92$\pm$0.07 & -1.20 & 0.84 & 1.95 & dI/BCD \\
\enddata
\tablenotetext{a}{For WLM and NGC 6822, distances for all fields were fixed at the values from \citet{albers19} and \citet{fusco12} respectively.}
\tablecomments{Columns: (1) Target galaxy; (2) Fraction of area out to 4.4 disk scalelengths covered by observations; (3) Foreground extinction \citep{sf11}; (4) Heliocentric distance in Mpc from the Extragalactic Distance Database \citep{tully09}; (5) Stellar Mass \citep{k18}; (6) HI mass \citep{k18}; (7) Abdundances from the \citet{marble10,berg12} compilations where available, with additional individual values from \citet{masegosa94,ks97,izotov97,hidalgogamez01,lgh03,vanzee06,garciabenito12,garciarojas16}.  (8) Tidal index $\theta_{1}$ \citep{k18}; (9) Distance to the nearest neighbor galaxy D(NN); (10) Distance to the nearest galaxy with Log M$_{\star}$/M$_{\odot}$$>$10; (11) Galaxy morphological type \citep{m12,k13}.}
\end{deluxetable}

\subsection{Observations and Photometry \label{photsect}}

The archival imaging of all of our target galaxies was obtained using one of the workhorse imagers onboard HST, either ACS/WFC (The Wide-Field Channel of the Advanced Camera for Surveys; \citealt{ford98}), WFC3/UVIS (the ultraviolet channel of Wide Field Camera 3; \citealt{kimble08}), or WFPC2 (Wide Field Planetary Camera 2; \citealt{holtzman95}).  The specific filters and instruments used vary across our sample, although all of our targets were imaged in the F814W filter of their respective camera as the redder of the two filters we use.  For the majority of our targets (34/42; 81\%), only a single pointing was available, while in some cases, multiple overlapping pointings were combined or, for the four targets with the largest projected angular sizes, up to 17 non-contiguous pointings were analyzed together (see Sect.~\ref{speccasesect}).  The observations we use for each target galaxy are summarized in Table \ref{obstab}.  

For pointings observed with ACS/WFC or WFC3/UVIS, we downloaded individual \texttt{flc} images, which have undergone standard pipeline preprocessing, including corrections for charge transfer inefficiency.  For WFPC2 pointings, we used individual \texttt{C0M} science images and the corresponding \texttt{C1M} data quality images.  We then used the 
\texttt{drizzlepac v3.0} package \citep{drizzlepac1,drizzlepac2} to construct a deep, distortion-corrected, drizzled reference image in each filter from the individual science images.  We performed additional preprocessing and point-spread function fitting (PSF) photometry using the software code \texttt{Dolphot} \citep{dolphot,dolphot2}, which uses model point-spread functions (PSFs) customized to each filter of each camera.  Briefly, preprocessing for each image includes applying a pixel area map, masking bad pixels, splitting into individual chips, and calculating a preliminary sky frame.  Details of how \texttt{Dolphot} performs PSF photometry can be controlled using various input parameters, and we use the parameters recommended by \citet{williams_phat,williams_phatter}\footnote{We retained \texttt{FitSky=3} to optimize photometric quality in crowded fields \citep{dalcanton09} and the corresponding aperture radius of \texttt{RAper}=8 for ACS/WFC and WFC3/UVIS recommended by the DOLPHOT manual.}, which have been applied to numerous recent imaging studies of Local Volume galaxies \citep[e.g.,][]{martin17,albers19,hargis20,collins22,mcquinn23,savino23,tran23,cohen24a,cohen24b,newman24}. 

\startlongtable
\begin{deluxetable}{lcccccccl}
\tabletypesize{\scriptsize}
\tablecaption{Observed Fields \label{obstab}}
\tablehead{
\colhead{Galaxy} & \colhead{Camera} & \colhead{Blue Filter} & \colhead{t$_{\exp}$ (blue) [s]} & \colhead{t$_{exp}$ (red) [s]} & \colhead{PID} & \colhead{C$_{\rm 50, blue}$} & \colhead{C$_{\rm 50, F814W}$} & \colhead{Archive Name}}
\startdata
WLM  &  ACS/WFC  &  F475W  &  27360  &  34050  &  13768 & 28.84,29.49 & 27.80,28.47 & NAME-WLM-GALAXY \\
     &  WFC3/UVIS  &  F475W  &  27360  &  34050  &  13768 & 29.33,29.43 & 28.22,28.30 & ANY \\
     &  WFPC2  &  F555W  &  3600  &  4700  &  6798 & 25.80,26.04 & 24.62,24.87 & \\
     &  WFPC2  &  F555W  &  5300  &  5400  &  6813 & 25.89,26.53 & 24.71,25.34 & WLM-STARCLUS \\
     &  ACS/WFC  &  F606W  &  2325  &  2187  &  15275 & 28.30,28.29 & 27.32,27.30 & WLM-GALAXY-POS1 \\
     &  ACS/WFC  &  F606W  &  2324  &  2185  &  15275 & 28.27,28.27 & 27.28,27.27 & WLM-GALAXY-POS2 \\
ESO410-005 & ACS/WFC & F606W & 13440 & 25200 & 10503 & 27.87,28.95 & 26.84,28.06 & \\
ESO294-010 & ACS/WFC & F606W & 13920 & 27840 & 10503 & 27.81,29.04 & 26.74,28.08 & \\
ESO540-032 & ACS/WFC & F606W & 11200 & 10062 & 10503 & 28.61,28.88 & 27.60,27.87 & \\
LGS3 & ACS/WFC & F475W & 15072 & 13824 & 10505 & 29.38,29.48 & 28.33,28.43 & \\
IC1613  &  ACS/WFC  &  F475W  &  31489  &  27119  & 10505 & 29.28,29.56 & 28.21,28.5 & \\
        &  ACS/WFC  &  F606W  & 6036 & 6036 & 13691 & 28.64,28.63 & 27.65,27.63\tablenotemark{a} & IC-1613-FIELD1 \\
        &  ACS/WFC  &  F606W  & 5952 & 5952 & 13691 & 28.64,28.63 & 27.65,27.63\tablenotemark{a} & IC-1613-FIELD2 \\
        &  WFPC2  &  F555W  &  19200  &  37200  &  7496 & 26.58,26.56 & 25.55,25.49 & \\
        &  WFPC2  &  F555W  &  10700  &  10700  &  6865 & 25.70,25.87 & 24.57,24.71 & \\
UGC685 & ACS/WFC  &  F606W  &  934  &  1226  &  10210 & 26.97,27.68 & 25.98,26.85 & UGC00685 \\
UGC1281  & ACS/WFC  & F606W &  937  &  1230  & 10210 & 26.88,27.66 & 25.82,26.78 & \\
Phoenix & ACS/WFC & F606W & 9347 & 9307 & 14734 & 28.64,28.93 & 27.65,27.96 & PHOENIX-I \\
        & WFC3/UVIS & F606W & 10087 & 10087 & 14734 & 28.87,28.87 & 27.78,27.77 & ANY \\
NGC0784 & ACS/WFC  &  F606W  &  933  &  1226  &  10210 & 26.86,27.69 & 25.85,26.82 & NGC0784\\
NGC2366 & ACS/WFC & F555W & 9560 & 9560 & 10605 & 27.54,28.33 & 26.47,27.53\tablenotemark{a} & NGC-2366-1 \\
        & ACS/WFC & F555W & 9560 & 9560 & 10605 & 27.54,28.33 & 26.47,27.53\tablenotemark{a} & NGC-2366-2 \\
UGC4459  &  ACS/WFC  &  F555W  &  4768  &  4768  &  10605  & 27.99,28.33 & 27.12,27.56 & UGC-04459 \\
UGC4483  &  WFC3/UVIS  &  F606W  &  36120  &  51600  & 15194 & 27.70,29.34 & 26.74,28.26 & \\
UGC5139  &  ACS/WFC  &  F555W  &  5829  &  5936  &  10605 & 28.35,28.56 & 27.50,27.78 & \\
UGC5364  &  WFC3/UVIS  &  F475W  &  19200  &  19520  &  10590 & 29.06,29.38 & 28.03,28.37 & LEOA-CENTER \\
SextansB  &  WFPC2  &  F606W  &  2700  &  3900  &  10915  & 25.46,26.44 & 24.37,25.35 & SEXB \\
NGC3109 & WFPC2 &  F606W  & 2700  &  3900  & 11307 & 25.15,26.34 & 24.08,25.19\tablenotemark{a} & NGC3109-WIDE1 \\
        &  & & 2700  &  3900  & 11307 & 25.15,26.34 & 24.08,25.19\tablenotemark{a} & NGC3109-WIDE2 \\
        &  &   & 2400  &  2400  & 11307 & 25.15,26.34 & 24.08,25.19\tablenotemark{a} & NGC3109-WIDE3 \\
        &  &    & 2400  &  2400  & 11307 & 25.15,26.34 & 24.08,25.19\tablenotemark{a} & NGC3109-WIDE4 \\
Antlia & ACS/WFC & F606W & 985 & 1174 & 10210 & 27.80,27.87 & 26.89,26.97 & \\
SextansA & WFPC2 & F555W & 1800 & 1800 & 5915 & 24.54,24.61 & 23.40,23.47 & DDO75 \\
         & WFPC2 & F555W & 19200 & 38400 & 7496 & 25.95,26.37 & 24.80,25.19 & DDO75 \\
NGC3738 & ACS/WFC  &  F606W  &  900  &  900  & 12546 & 26.39,27.79 & 25.31,26.82 & \\
NGC3741  &  ACS/WFC  &  F475W  &  2262  &  2331  & 10915 & 27.13,28.26 & 26.01,27.35 & \\
UGC6817  &  WFPC2  &  F606W  &  4800  &  7200  & 11986 & 26.15,27.15 & 25.05,26.04 & \\
NGC4068 & ACS/WFC & F606W  &  1200  &  900  & 9771 & 27.16,27.83 & 26.15,26.79 & \\
NGC4163  &  ACS/WFC &  F606W  & 1200  &   900  &   9771 &  26.91,28.41 & 25.84,27.40\tablenotemark{a} & \\  
         &          &         & 2292  &  2250  &  10915 & 26.91,28.41 & 25.84,27.40\tablenotemark{a} &  \\
NGC4190  &  WFPC2  &  F606W  &  2200  &  2200  &  10905 & 26.91,28.41 & 25.84,27.40\tablenotemark{a} & NGC-4190 \\
UGCA281 & ACS/WFC & F606W & 1148 & 938 & 10905 & 25.69,27.84 & 24.74,26.96 & \\
UGC7577  &  WFPC2  &  F606W  &  4800  &  7200  &  11986 & 25.88,26.74 & 24.78,25.59 & \\
UGCA292  &  ACS/WFC  &  F475W  &  2250  &  2274  &  10915 & 28.28,28.33 & 27.34,27.40 & \\
UGC8024  &  ACS/WFC  &  F606W  &  924  &  1128  &  10905 &  27.47,27.77 & 26.58,26.89 & \\
UGC8091  &  ACS/WFC  &  F475W  &  2244  &  2259  &  10915 & 27.94,28.32 & 26.97,27.37 & \\
UGCA320 & ACS/WFC  & F606W &  1030 & 1030 & 14636 & 27.02,27.74 & 26.04,26.77 & \\
UGC8201 & ACS/WFC  &  F555W  &  4768  &  4768  & 10605 & 27.85,28.39 & 26.91,27.63 & \\
UGC8508  &  ACS/WFC  &  F475W  &  2280  &  2349  &  10915 & 27.30,28.29 & 26.20,27.39 & \\
UGC8638 &  ACS/WFC  &  F606W  &  1200  &  900  &  9771 & 27.23,27.87 & 26.22,26.79 \\
UGC8651  &  ACS/WFC  &  F606W  &  1016  &  1209  &  10210 & 27.52,27.87 & 26.58,26.98 & \\
UGC8837  & ACS/WFC  & F606W & 946  &  1158  &  10905 & 27.44,27.87 & 26.52,27.00 & \\
UGC9128  &  ACS/WFC  &  F606W  &   985  &  1174  &  10210 & 27.55,27.93 & 26.59,27.05 & \\
UGC9240  &  ACS/WFC  &  F606W  &  2301  &  2265  &  10915 & 27.16,28.40 & 26.10,27.43 & \\
IC4662  &  ACS/WFC  &  F606W  &  1200  &  900  &  9771 & 25.43,27.50 & 24.38,26.47 & \\
NGC6822 &  ACS/WFC  &  F606W  &  8660  &  8350  &  14191 & 28.54,28.58 & 27.49,27.56 & NGC6822.BAR.EAST \\
  &  ACS/WFC  &  F475W  &  886  &  1346  &  12180 & 27.40,27.43 & 26.54,26.56 & NGC6822-GRID1 \\
  &  ACS/WFC  &  F475W  &  1119  &  1118  &  12180 & 27.53,27.59 & 26.56,26.61 & NGC6822-GRID2 \\
  &  ACS/WFC  &  F475W  &  1119  &  1118  &  12180 & 27.54,27.55 & 26.57,26.58 & NGC6822-GRID3 \\
 &  ACS/WFC  &  F475W  &  1119  &  1118  &  12180 & 27.36,27.54 & 26.32,26.54 & NGC6822-GRID4 \\
  &  ACS/WFC  &  F606W  &  1554  &  2271  &  15336 & 26.68,27.12 & 25.59,26.08 & NGC6822-HUBBLEVII  \\
  &  WFC3/UVIS  &  F606W  &  2411  &  5062  &  15336 & 27.54,27.56 & 26.64,26.68 & ANY  \\
&  ACS/WFC  &  F606W  &  2331  &  2469  &  15336 & 28.18,28.21 & 27.23,27.27 & NGC6822-SC2  \\
&  WFC3/UVIS  &  F606W  &  2411  &  2531  &  15336 & 28.10,28.11 & 26.99,27.00 & ANY  \\
 &  ACS/WFC  &  F606W  &  2331  &  2469  &  15336 & 27.95,27.95 & 27.02,27.00 & NGC6822-SC3 \\
 &  WFC3/UVIS  &  F606W  &  2411  &  2531  &  15336 & 27.92,27.93 & 26.84,26.85 & ANY  \\
&  ACS/WFC  &  F606W  &  2331  &  2469  &  15336 & 27.92,27.93 & 26.99,27.00 & NGC6822-SC5  \\
&  WFC3/UVIS  &  F606W  &  2411  &  2531  &  15336 & 27.79,27.83 & 26.71,26.75 & ANY  \\
 &  ACS/WFC  &  F606W  &  2331  &  2469  &  15336 & 27.95,27.95 & 27.00,27.02 & NGC6822-SC6 \\
 &  WFC3/UVIS  &  F606W  &  2411  &  2531  &  15336 & 27.86,27.86 & 26.79,26.79 & ANY  \\
 &  ACS/WFC  &  F606W  &  2331  &  2469  &  15336 & 27.99,27.98 & 27.05,27.03 & NGC6822-SC7 \\
 &  WFC3/UVIS  &  F606W  &  2411  &  2531  &  15336 & 27.89,27.86 & 26.80,26.77 & ANY  \\ 
DDO210 & ACS/WFC & F475W & 22910 & 33480 & 12925 & 28.58,28.94 & 27.54,27.87 & \\
IC5152  &  WFPC2  &  F606W  &  4800  &  9600  &  11986 & 24.49,26.15 & 23.36,25.02 & I5152 \\
\enddata
\tablenotetext{a}{Contiguous pointings of similar depth observed with the same instrument and filter combination (typically within the same program) were combined for SFH analysis.}
\tablecomments{For targets with multiple pointings, each pointing is listed on a separate line.  Columns: (1) Target name (2) Instrument and camera (3) The bluer of the two filters used for imaging; in all cases the redder filter is F814W of the respective camera. (4) Total exposure time in seconds in the bluer filter (5) Total exposure time in the F814W filter (6) Program ID (7) 50\% faint completeness limit in the bluer of the two filters, listed in the third column. (8) 50\% completeness in the redder of the two filters, which is F814W of the respective instrument in all cases. (9) Target name in the MAST archive when different from the name listed in the first column.}
\end{deluxetable}

The raw photometric catalog output by \texttt{Dolphot} is in the Vegamag system, and has several diagnostic parameters which are useful to eliminate poorly measured, non-stellar and/or spurious detections.  These parameters are provided for each star as global values, per-filter values, and per-image values.  
We impose the following cuts on our photometric catalogs using the per-filter diagnostic parameters to retain only well-measured stellar sources:

\begin{enumerate}
    \item Object type $\leq$2.  This choice eliminates sources that are non-stellar based on intensity distributions that differ from the model PSF (e.g., elongated, extended or overly concentrated).
    \item Signal-to-noise ratio SNR $>$4 in each filter.
    \item Photometric quality flag $<$4 in each filter.  This is a bitwise flag used to eliminate stars that have too many bad or saturated pixels to obtain a meaningful PSF fit.  For example, our chosen value, required in each filter separately, will reject sources with saturated cores as well as those with a large fraction of their photometry aperture extending off of a detector.
    \item The \texttt{sharp} parameter measures whether a source is more or less centrally concentrated than the model PSF, with positive values representing more concentrated sources (i.e., cosmic rays) and negative values representing more extended sources (i.e., blends or background galaxies).  We retain sources with $({\rm sharp}_{1} + {\rm sharp}_{2})^{2}$ $<$ 0.075, where the subscripts denote each of the two filters.
    \item The \texttt{crowd} parameter measures how much brighter (in magnitudes) a star would have been if its neighbors had not been fit with a PSF and subtracted.  We require $({\rm crowd}_{1} + {\rm crowd}_{2})$ $<$ 0.1.
\end{enumerate}

The procedure of calculating the best-fit SFH for a particular observed CMD relies on our ability to apply the noise properties of our data to synthetic photometry.  Therefore, for each target galaxy, we insert $>$10$^{6}$ artifical stars with a color-magnitude distribution sufficiently broad to sample the full distribution of the observed stellar populations.    
Artificial stars are considered recovered if they pass all of the photometric quality cuts listed above that are applied to real sources and have output magnitudes within 2 mag of their input magnitudes (to minimize false positives, e.g., \citealt{paust09}). 

\section{Measuring Radial Age Gradients \label{methodsect}}

The radial age gradient in each galaxy (and its uncertainty) is quantified via three steps, 
illustrated for an example case in Fig.~\ref{sfh_example_fig}.  These three steps are summarized here before providing further details on each analysis step in turn:

\begin{enumerate}
\item Each galaxy is divided spatially into four equally populated concentric elliptical annuli (see Sect.~\ref{structuralsect}), which we refer to as \enquote{radial bins} (Fig.~\ref{sfh_example_fig}a).  

\item We fit a SFH to the CMD of the stars located in each radial bin (Fig.~\ref{sfh_example_fig}b, see Sect.~\ref{sfhsect}), yielding the lookback time by which 90\% and 50\% of the cumulative stellar mass was formed in that radial bin, denoted $\tau_{90}$ and $\tau_{50}$ respectively (see Fig.~\ref{sfh_example_fig}d). 

\item The radial age gradients $\gamma_{90}$ and $\gamma_{50}$ are the gradient in $\tau_{90}$ and $\tau_{50}$ versus semimajor-axis-equivalent distance from the center of the galaxy, normalized to the half-light radius R$_{hl}$ (lower right panels of Fig.~\ref{sfh_example_fig}).    
\end{enumerate}
Details of each of these steps are as follows:  

\begin{figure}
\gridline{\fig{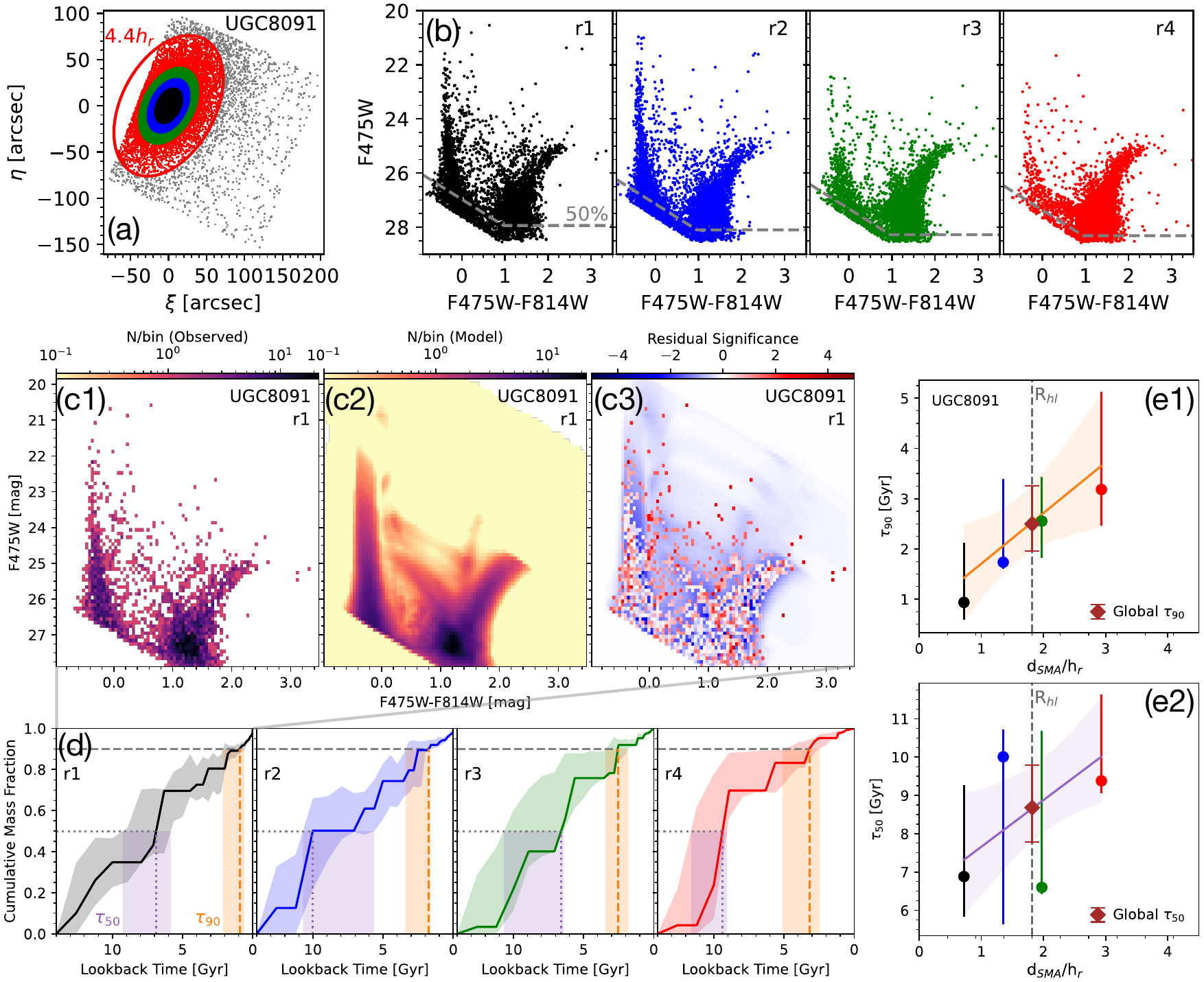}{0.99\textwidth}{}}
\caption{An example of our procedure to determine radial gradients in $\tau_{50}$ and $\tau_{90}$ for a target galaxy.  \textbf{Top Row:} A spatial density map of observed sources is shown in panel (a) with the boundaries of the four equally populated elliptical annuli (\enquote{radial bins}) overplotted.  CMDs of the sources in each radial bin are shown in panel (b), color-coded as in (a).  The 50\% completeness limits, used as the faint limits of the portion of the CMD used for SFH fitting, are indicated using dashed grey lines.  The completeness limits become fainter due to reduced crowding in radial bins farther from the galaxy center.  \textbf{Center:} Panels (c1)-(c3) illustrate an example SFH fit to the innermost radial bin (denoted r1, shown in black in panels (a) and (b)), showing Hess diagrams (i.e., two-dimensional color-magnitude density plots) of the observed (c1), modeled (c2), and residual (observed$-$modeled) source densities, with the latter in units of Poisson standard deviations (c3).  \textbf{Bottom Row:} Panel (d) shows CSFHs fit to each of the four radial bins in panels (a) and (b), illustrating the values and uncertainties of $\tau_{50}$ and $\tau_{90}$ for each, calculated via interpolation in the CSFH and its uncertainty envelope.  \textbf{Lower Right:} Radial gradients in $\tau_{90}$ (e1) and $\tau_{50}$ (e2), denoted $\gamma_{90}$ and $\gamma_{50}$ respectively, resulting from maximum likelihood fits to the per-radial-bin values.  The radial gradient slopes are normalized to the half-light radius R$_{hl}$, shown as a dashed vertical line.  The values of $\tau_{90}$(global) and $\tau_{50}$(global) are overplotted as diamonds, calculated by interpolating the linear fits (and uncertainties) at R$_{hl}$ following \citet{graus19}.
\label{sfh_example_fig}}
\end{figure}

\subsection{Structural Parameters \label{structuralsect}}

To measure radial age gradients in our target galaxies, each galaxy is divided spatially into four radial bins (in practice, concentric elliptical annuli).  While a linear fit (i.e., with two free parameters) could, in a strict sense, be achieved with only three radial bins, we found that the use of four radial bins improved the stability of the fits.  Functionally, the choice of four radial bins 
is a necessary compromise between spatial resolution (needed to measure radial age gradients within each galaxy) and number statistics (which drive statistical uncertainties on SFH fits).  While some target galaxies hosted a sufficiently large number of well-detected sources that more than four radial bins could have been used, we chose to fix the number of radial bins per target galaxy to keep our analysis as self-consistent as possible, while we demonstrate in Appendix~\ref{testsect} that the use of more finely divided radial bins does not significantly affect measured age gradients.  To delineate the radial bins, the per-galaxy center, position angle $\theta$, and minor-to-major axis ratio $b/a$ were calculated using ellipse fits to \textit{Spitzer} IRAC 3.6$\mu$ imaging, meticulously cleaned of foreground contaminants, detailed in McQuinn et al.~(in prep.).  With the location of the galaxy center and the ellipse parameters $\theta$ and $b/a$ in hand, a semi-major-axis-equivalent distance $d_{\rm SMA}$ from the center of a galaxy can then be calculated for any star with tangent plane coordinates $\xi$ and $\eta$ relative to the galaxy center.  If we set:

\begin{equation}
u = \xi \cos \theta - \eta \sin \theta
\end{equation}
\begin{equation}
v = \xi \sin \theta + \eta \cos \theta
\end{equation}
then:
\begin{equation}
d_{\rm SMA} = \sqrt{\frac{u^{2}+v^{2}}{(b/a)^{2}}}
\end{equation}

We opted to fix the outer boundary of the outermost radial bin at $d_{\rm SMA}$=4.4$h_{r}$ for all galaxies, where $h_{r}$ is the disk scalelength from exponential surface brightness profile fits to the deprojected 3.6$\mu$ fluxes.  The use of a fixed outer boundary for each galaxy (in terms of its $h_{r}$) is necessary to keep our analysis self-consistent across our entire sample given galaxy-to-galaxy differences in spatial coverage (due, for example, to varying physical size, distance, axis ratios and field of view of archival observations).  The choice of 4.4$h_{r}$ for the location of the outermost ellipse boundary is a necessary compromise between maximizing the number of target galaxy sources included while retaining as many target galaxies as possible with imaging coverage out to the chosen boundary, and circumscribes $\sim$93.4\% of the galaxy light assuming an exponential profile.\footnote{\citet{leroy21} find that star-forming galaxies are somewhat more concentrated than pure exponentials, with $R_{hl}$/$h_{r}=$1.41 compared to $R_{hl}$/$h_{r}=$1.68 for a pure exponential.  Across our sample, we find a mean $R_{hl}$/$h_{r}=$1.63 with a standard deviation of 0.28 excluding cases where $R_{hl}$ and $h_{r}$ were not measured independently.}  
Within 4.4$h_{r}$, we sort stars by their $d_{\rm SMA}$ to divide the four ellipses such that each ellipse contains an equal number of observed stars, with the goal of obtaining similar SFH uncertainties in each ellipse.  

Lastly, a few target-specific caveats: For UGCA292 and UGC5364 (=Leo A), structural parameters were measured from $R$-band imaging, due to very low surface brightness and corruption respectively in the 3.6$\mu$ imaging.  In addition, a minority of our targets (12/42; 29\%) did not have structural parameters measured from 3.6$\mu$ imaging in the McQuinn et al.~(in prep.) study.  We therefore selected structural parameters from the literature attempting to maintain as much homogeneity in our analysis as possible.  To this end, where available (4 targets), R$_{hl}$ and $h_{r}$ were drawn from values similarly measured from 3.6$\micron$ Spitzer imaging by \citet{lelli16}\footnote{Comparing R$_{hl}$ and $h_{r}$ value for the 10 targets in common between our study and \citet{lelli16}, agreement is generally good, with the \citet{lelli16} values of R$_{hl}$ running on average 18$\pm$5\% smaller ($\sigma$=16\%) and $h_{r}$ values 4$\pm$9\% larger ($\sigma$=29\%).}.  For these targets, we used position angles and axis ratios from the HyperLEDA database\footnote{\url{http://atlas.obs-hp.fr/hyperleda/}} \citep{makarov14}, producing ellipse fits in good agreement with the observed stellar distributions.  For one additional target (ESO540-032), structural parameters were only available from \citet{kirby08}, and for the remainder of the targets, we used structural parameters from the \citet{m12} compilation, modulo updates from deep, wide-field ground-based studies of Phoenix, IC 1613 and NGC 6822 \citep{battaglia12,bernard07,pucha19,tantalo22}.  
    
\subsection{Star Formation History Fitting \label{sfhsect}}

In each radial bin, we use the software code \texttt{MATCH} \citep{match} to calculate the best-fit SFH.   
\texttt{MATCH} has been used to measure the SFHs of scores of galaxies throughout the Local Volume over their entire lifetimes \citep[e.g.,][]{williams09,mcquinn10b,weisz11,williams11,radburnsmith12,weisz14,geha15,williams17,mcquinn18,savino23,mcquinn23} and produces results comparable with other CMD synthesis codes \citep{monelli10a,monelli10b,hidalgo11,skillman14,skillman17,garling24}.
\texttt{MATCH} uses the well-known CMD synthesis technique (see \citealt{tolstoy09} for a review) to forward model the observed CMD of each radial bin as a linear combination of simple stellar populations, seeking the best fit based on a Poisson maximum likelihood statistic.  

Operationally, \texttt{MATCH} calculates the star formation rate and metallicity in user-selected time intervals under a set of input assumptions, and our chosen input values are based on the vast database of previous SFH studies of Local Volume dwarfs using \texttt{MATCH} \citep[e.g.,][]{weisz13,weisz14,savino23,mcquinn24a,mcquinn24b}.  Specifically, we use time bins covering lookback times of 6.6$\leq$Log$_{\rm 10}$(Age/yr)$\leq$10.15, with width $\delta$Log$_{\rm 10}$(Age/yr)$=$0.1 for lookback times of Log$_{\rm 10}$(Age/yr)$\leq$9 and $\delta$Log$_{\rm 10}$(Age/yr)$=$0.05 otherwise. 
We assume a \citet{kroupa} stellar initial mass function (IMF) and a stellar binary fraction of 0.35 with a flat mass ratio distribution (see, e.g., \citealt{cohen24b} for further discussion of this choice).   
We impose a loose astrophysically motivated prior on the age-metallicity relation (AMR), requiring that the metallicity increases monotonically with time, from an initial value of $-$2.1$\leq$[M/H]$\leq$$-$1.0 dex to a present-day value of $-$1.5$\leq$[M/H]$\leq$0.2 dex, with a spread of $\Delta$[M/H]$=$0.15 dex in each time bin.  This loose prior helps to break the age-metallicity degeneracy in SFH fitting since most of our imaging is too shallow to reach the ancient ($\sim$13 Gyr) main sequence turnoff.  
 
To infer the best-fit SFH, \texttt{MATCH} compares binned observed and synthetic CMDs, with a user-specified resolution of 0.05 mag on the color (horizontal) axis and 0.1 mag on the magnitude (vertical) axis.  For consistency across our sample, we restrict the region of the CMD used for SFH fitting to lie brightward of the 50\% faint completeness limit in each radial bin.  The use of completeness limits calculated on a per-bin basis exploits the fainter completeness limits in the less crowded outer radial bins of our target galaxies, and counteracts the tendency of the outer radial bins to cover a disproportionately large fraction of the studied range of d$_{\rm SMA}$ due to stellar radial density profiles that decrease (approximately exponentially) with increasing d$_{\rm SMA}$ when designating equally-populated radial bins.  Our observations span a range of photometric depth, and assuming the distances and extinctions from Table \ref{obstab} to convert F814W values to an absolute scale, their innermost (outermost) ellipses have a median depth (50\% faint completeness limit) of M$_{\rm F814W}$=-0.3 (0.2) mag across our sample and interquartile ranges of 4.4 (3.4) mag.  Half of our pointings have F814W 50\% completeness limits deeper than the red clump (assuming M$_{\rm F814W,RC}\sim$$-$0.3; e.g., \citealt{parsec}) for their innermost ellipses, and 65\% for their outermost ellipses.  The 50\% faint completeness limits of the innermost and outermost ellipses are provided for all of our observations in both filters in Table \ref{obstab}.

When inferring the SFH of a radial bin, we allow target distance modulus $(m-M)_{0}$, foreground extinction A$_{V,fg}$ and internal differential extinction $\delta$A$_{V}$ to vary, imposing data-driven priors.  We first perform a SFH fit to the outermost of the four radial bins in each target galaxy, which has the deepest completeness limit in all cases (recall that all ellipses have an equal number of observed sources), allowing all three of these quantities to vary by $\pm$0.1 mag in increments of 0.05 mag around the distance modulus obtained from the tip of the red giant branch (TRGB) provided in the Extragalactic Distance Database \citep[EDD;][]{tully09} and the foreground extinction from the \citet{sf11} recalibration of the \citet{sfd98} dust emission maps.  Such a grid-search approach to optimizing SFH fits with \texttt{MATCH} has been tested and applied previously, yielding excellent results versus independent values where available \citep{lewis15,lazzarini22,cohen24a,cohen24b}.  The allowed range of $\pm$0.1 mag in $(m-M)_{0}$ and A$_{V,fg}$ are based on reported uncertainties for the target galaxy distances (see Table \ref{propstab}) and extinctions \citep{sf11}, and we demonstrate in  Appendix~\ref{testsect} that our radial age gradient slopes are not significantly affected by changing assumptions on $(m-M)_{0}$ and A$_{V,fg}$ at this level.  Once the best-fit $(m-M)_{0}$ and A$_{V,fg}$ are obtained from the outermost radial bin, these values are fixed for the three remaining ellipses\footnote{The spatial resolution of the \citet{sfd98} maps, set by the beam size of 6.1$\arcmin$ FWHM in the far-infrared dust emission imaging, does not exceed the spatial coverage of our targets except for four special cases described below, in which foreground extinction was set on a per-field basis (see Sect.~\ref{speccasesect}).}, while internal differential extinction $\delta$A$_{V}$ is allowed to float on a per-radial-bin basis.  While late-type dwarfs generally harbor little internal extinction \citep[e.g.,][]{mcquinn15,kahre18}, exceptions do exist, such as the SMC \citep[e.g.,][]{skowron21}.  

To directly examine the impact of different stellar evolutionary libraries assumed for SFH fitting, we perform our SFH fitting procedure twice for each radial bin of each target galaxy, using both the PARSEC \citep{parsec} and MIST \citep{mist1,mist2} libraries of solar-scaled stellar evolutionary models.  For each of these assumed libraries, we provide the best-fit values of $(m-M)_{0}$ and A$_{V,fg}$ for our target galaxies in Table \ref{slopetab}.  Uncertainties on our SFHs include both a random (statistical) and systematic component, calculated using prescriptions detailed in \citet{dolphin_randerr} and \citet{dolphin_syserr} respectively.

\subsection{Calculating Radial Age Gradients \label{calcsect}}

We characterize radial stellar age gradients in our target galaxies using the cumulative star formation history (CSFH) output by \texttt{MATCH}, which quantifies the fraction of cumulative stellar mass ever formed as a function of lookback time.  The CSFH is advantageous for measuring radial stellar age gradients because it is, by definition, a cumulative distribution, and is therefore sensitive to the shape of the lifetime SFH and insensitive to its normalization (which can depend, for example, on the assumed IMF).  To measure radial age gradients, we interpolate in the CSFH for each radial bin to calculate the lookback times to form 50\% and 90\% of the cumulative stellar mass, 
denoted $\tau_{50}$ and $\tau_{90}$ respectively.  The uncertainties on $\tau_{50}$ and $\tau_{90}$ are a direct result of interpolating in the CSFH uncertainties provided by \texttt{MATCH}, which correspond to 16\% and 84\% confidence intervals and can be asymmetric.  This interpolation procedure is illustrated for all four radial ellipses corresponding to an example target galaxy in Fig.~\ref{sfh_example_fig}d, where the CSFH is shown in black, its 1-$\sigma$ (16th to 84th percentile) confidence interval is shown in grey, and the values of $\tau_{50}$ and $\tau_{90}$ are shown using vertical lines with shading representing their resulting asymmetric uncertainties.

The radial age gradients $\gamma_{50}$ and $\gamma_{90}$ are the gradient in per-radial-bin $\tau_{50}$ and $\tau_{90}$ versus d$_{\rm SMA}$ (see the right-hand panels of Fig.~\ref{sfh_example_fig}), normalized to the half-light radius of each galaxy.  
We provide the best-fit values of $\gamma_{50}$ and $\gamma_{90}$ from posterior distributions calculated using a Markov Chain Monte Carlo (MCMC) approach as implemented using the \texttt{emcee} package \citep{emcee}.  
We assume a split-Gaussian likelihood function, accounting for highly asymmetric uncertainties in per-radial-bin $\tau_{50}$ or $\tau_{90}$, and also make the necessary and conservative assumption that bin-to-bin uncertainties in $\tau_{50}$ or $\tau_{90}$ are uncorrelated.
For each target galaxy, 50 walkers were run for 500 burnin iterations followed by 3000 production iterations (sufficient given autocorrelation lengths of $<$30 iterations in all cases), and the uncertainties we report correspond to the 16th and 84th percentiles of the posterior distributions of the slopes from the linear fits.  The individual age gradients fits for each of our target galaxies are illustrated in Appendix~\ref{indslopesect}.  

We also provide \enquote{global} galaxy-wide values of the lookback times $\tau_{50}$ and $\tau_{90}$ denoted $\tau_{50}$(global) and $\tau_{90}$(global).  These values are used to characterize the per-galaxy lifetime mass assembly histories of our targets, and we quantify the significance of their correlations versus $\gamma_{50}$ and $\gamma_{90}$ observationally in Sect.~\ref{obscorrsect} and compare them to quantitative predictions from multiple sets of simulations in Sect.~\ref{simsect}.  Since spatial coverage of some of our target galaxies is (azimuthally) incomplete (i.e., $\mathcal{A}$$<$1 in Table \ref{propstab}), we opt against combining the SFHs inferred from individual radial bins, instead calculating $\tau_{50}$(global) and $\tau_{90}$(global) using a technique that is agnostic to spatial coverage that may differ among targets as well as among individual radial bins within each target.  Specifically, we leverage the finding by \citet{graus19} that galaxy-wide values of $\tau_{50}$ and $\tau_{90}$ are equivalent to values calculated at R$_{hl}$.  The values of $\tau_{50}$(global) and $\tau_{90}$(global) that we present are therefore calculated by interpolating in the best-fit line (and 1-$\sigma$ uncertainties) from the maximum likelihood fit to the per-radial-bin values of $\tau_{50}$ and $\tau_{90}$ versus d$_{SMA}$, evaluated at R$_{hl}$.  Examples illustrating the global galaxy-wide $\tau_{50}$ and $\tau_{90}$ values calculated using this procedure are shown in brown in the lower right-hand panels of Fig.~\ref{sfh_example_fig}.  To validate the hypothesis that the procedure of interpolating our best-fit radial gradient at R$_{hl}$ is representative of galaxy-wide values of $\tau_{90}$ and $\tau_{50}$, we turn to the subset of our targets with complete spatial coverage out to at least 4.4$h_{r}$ (i.e., with $\mathcal{A}$=1 in Table \ref{propstab}).  Across this subset, our interpolated values $\tau_{50}$(global) and $\tau_{90}$(global) agree with directly measured values to within their 1-$\sigma$ uncertainties in the vast majority (86\%) of cases.  

\subsection{Special Cases \label{speccasesect}}

There are four Local Group dwarfs that have sufficiently large projected angular sizes that multiple non-contiguous imaging fields were available for SFH fitting.  These are WLM, NGC 6822, IC 1613, and Phoenix, and the individual pointings used for SFH fitting are listed in Table \ref{obstab}.  For each of these galaxies, we imposed the astrophysical prior that all fields for a given galaxy lie at the same distance by fixing the distance modulus for all radial bins of all fields to the EDD value (or updated values in the case of WLM and NGC 6822; see Table \ref{propstab}), and the foreground extinction of each field to the per-field value from the \citet{sf11} maps (internal differential extinction $\delta$A$_{V}$ was still allowed to float on a per-radial-bin basis as with all of our targets).  However, the choice to fix $(m-M)_{0}$ and A$_{V,fg}$ for these four galaxies did not ultimately impact our results at a statistically significant level, and allowing them to float on a per-field basis yielded radial stellar age gradients $\gamma_{90}$ and $\gamma_{50}$ and per-galaxy lookback times $\tau_{90}$(global) and $\tau_{50}$(global) that were unaffected to within their 1-$\sigma$ uncertainties.  While our choice to fit age gradients to all radial bins of all fields for each target (listed in Table \ref{obstab}) necessarily neglects the possibility of azimuithal variations in age gradients, this approach is consistent with our analysis for the remainder of our sample.  Furthermore, when a sufficient number of individual pointings are available, individual deviations from the best-fit galaxy-wide age gradients may become apparent, with pertinent astrophysical implications (e.g., \citealt{cohen24a,cohen24b,cohen25}, also see Sect.~\ref{litsect}).  Radial gradient slopes for these four galaxies are illustrated in the bottom row of each of the figures presented in Appendix~\ref{indslopesect}.  

\section{Results and Discussion \label{resultsect}}

Our best-fit radial age gradient slopes $\gamma_{50}$ and $\gamma_{90}$ are reported in Table 
\ref{slopetab} along with their 1-$\sigma$ uncertainties, preceded by the assumed structural parameters for each target galaxy (and their source).  Comparing results based on PARSEC versus MIST stellar evolutionary models, the MIST models generally predict slightly steeper radial age gradients, but not at high statistical significance: The difference $\Delta$$\gamma_{90}$ (MIST-PARSEC) has a (weighted) mean of 0.24$\pm$0.18 Gyr/R$_{hl}$ with a standard deviation of $\sigma$=0.53 Gyr/R$_{hl}$.  For $\gamma_{50}$, the difference in radial gradient slopes assuming PARSEC versus MIST stellar evolutionary models is insignificant ($\langle$$\Delta$$\gamma_{50}$$\rangle$=$-$0.04$\pm$0.30). 
We therefore illustrate our primary results in subsequent figures assuming PARSEC evolutionary models, although we provide quantitative results assuming MIST models as well to demonstrate the insensitivity of our results to the stellar evolutionary library assumed for SFH fitting.   

\subsection{Empirical Correlations With Global Galaxy Properties \label{obscorrsect}}

To gain insight into the astrophysical drivers behind our measured radial age gradient slopes, we searched for statistical correlations between each of the age gradient slopes $\gamma_{90}$ and $\gamma_{50}$ versus seven global galaxy properties:

\begin{itemize}
\item $\tau_{90}$(global) and $\tau_{50}$(global).  Simulations predict that radial gradient slopes are correlated with the overall global mass assembly histories of dwarfs (\citealt{graus19,riggs24}, also see Sect.~\ref{introsimsect}), and we return to this point in more detail below in Sect.~\ref{simsect}.

\item Gas-phase oxygen abundance 12+Log(O/H), taken from the compilations of \citet{marble10} and \citet{berg12} for the majority of our sample, plus additional values for individual galaxies absent from those compilations \citep{masegosa94,ks97,izotov97,hidalgogamez01,lgh03,vanzee06,garciabenito12,garciarojas16}.  
While spectroscopic metallicities based on individual member stars exist for some Local Volume dwarfs \citep[e.g.,][]{kirby13,taibi22}, they tend to be observationally biased towards the nearest (Local Group) targets, covering only a minority of our galaxy sample.  Conversely, gas-phase oxygen abundances are available for most (35/42) of our targets, while self-consistent metallicity measurements for numerous member stars within a large sample of Local Volume dwarfs may be forthcoming, for example, from integral field unit observations \citep{li25}).

\item Stellar masses and H\textsc{I}-to-stellar mass ratios from the \citet{k18} compilation.  

\item We also include three criteria quantifying the environments of our target galaxies.  These include the tidal index $\theta_{1}$ and D(NN), the distance to the nearest neighboring galaxy, regardless of mass.\footnote{We have not included other indicators of local density contrast (e.g., $\theta_{5}$) since they correlate strongly with $\theta_{1}$ \citep{k18}.}  Beyond nearby neighbors, we also wish to address whether large-scale environment (i.e., proximity to a massive host) has any correlation with age gradients, as predicted for \textit{global} SFHs by simulations \citep{christensen24}.  Therefore, we also test for correlations between our age gradient slopes versus the distance to the nearest \enquote{large} galaxy with  
M$_{\star}$$\geq$10$^{10}$M$_{\odot}$, denoted 
D(NLG$_{10}$).\footnote{We also tested lowering the threshold for \enquote{massive} galaxies to M$_{\star}$$\geq$10$^{9}$M$_{\odot}$, and found correlation coefficients consistent with those for D(NLG$_{10}$) to well within their uncertainties.}

\end{itemize}

\begin{figure}
\gridline{\fig{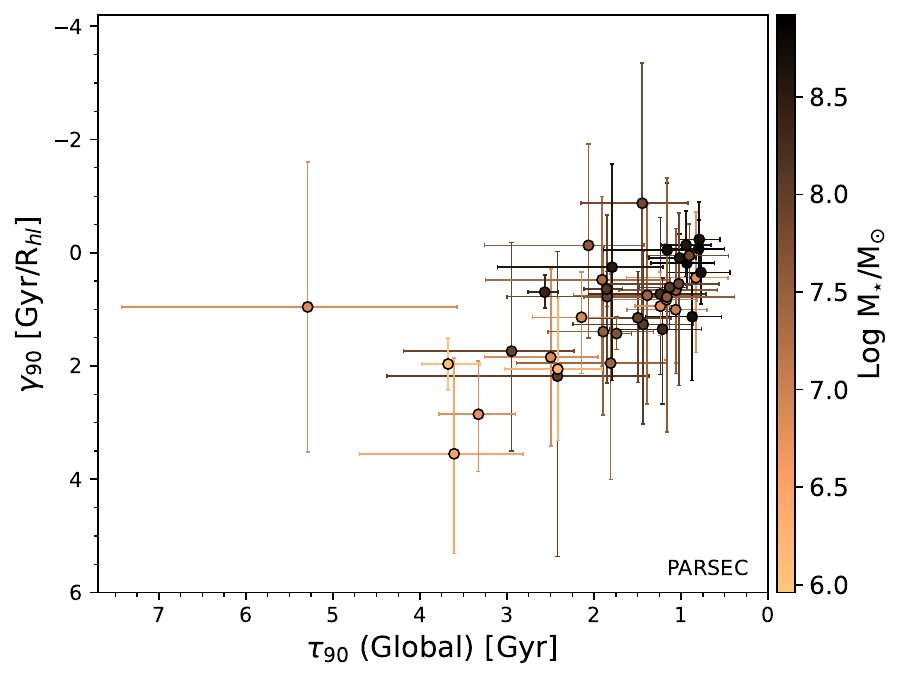}{0.49\textwidth}{}
          \fig{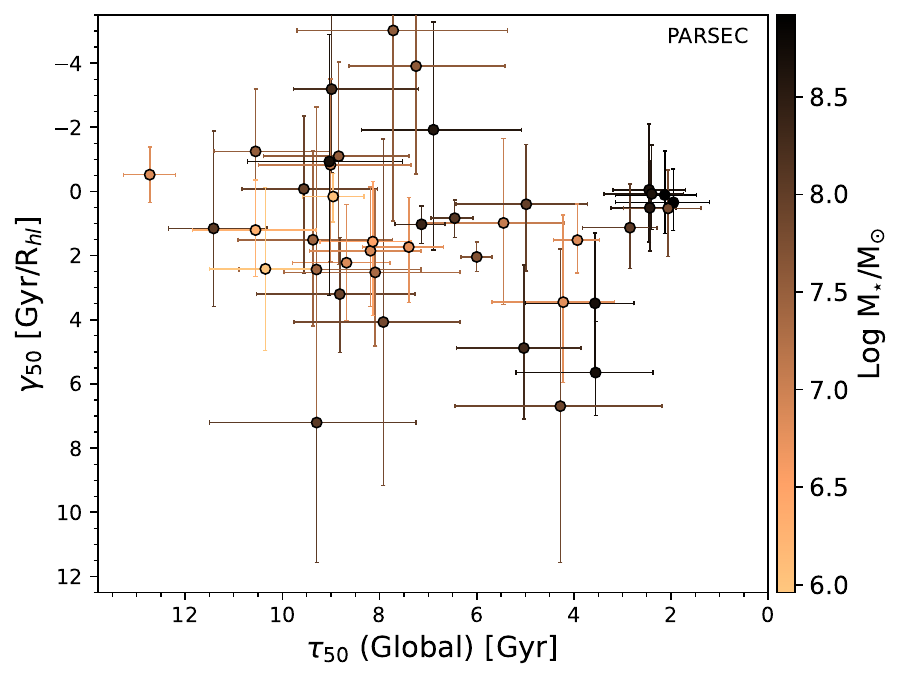}{0.49\textwidth}{}}
\caption{\textbf{Left: }Radial age gradient slopes $\gamma_{90}$ plotted versus galaxy-wide $\tau_{90}$(global) values for each of our target galaxies, color-coded by their stellar mass.  Results shown here assume PARSEC stellar evolutionary models for SFH fitting, but results assuming MIST models are also presented in Table \ref{slopetab}.  We detect a highly significant ($p$-values $\lesssim$0.001) linear correlation between $\gamma_{90}$ and $\tau_{90}$(global), and a moderately significant correlation with stellar mass such that lower-mass galaxies have steeper present-day outside-in radial gradients in $\tau_{90}$ (see text for details).  For clarity, axis limits are set to exclude one target with the earliest value of $\tau_{90}$(global) and steepest value of $\gamma_{90}$ (LGS3; see Sect.~\ref{t90sect}).  \textbf{Right: }Same, but for $\gamma_{50}$ versus $\tau_{50}$(global).  Unlike $\gamma_{90}$, we do not find any significant correlations between $\gamma_{50}$ and any of the global galaxy properties we examine, including $\tau_{50}$(global).  
\label{obstaufig}}
\end{figure}

To search for correlations between $\gamma_{90}$ and $\gamma_{50}$ versus each global parameter, we calculate the Pearson linear correlation coefficient $\rho$ and the Spearman rank correlation coefficient $r_{s}$ and their $p$-values.   
Uncertainties on the correlation coefficients and $p$-values correspond to the 16th and 84th percentiles resulting from 10,000 monte carlo draws from (asymmetric) Gaussian deviates with standard deviations equal to the 1-$\sigma$ uncertainties of each parameter.  Our search for correlations between $\gamma_{50}$ or $\gamma_{90}$ versus global galaxy properties reveals the following:

\begin{itemize}

\item The most statistically significant correlation, by far, occurs between the radial age gradient slope $\gamma_{90}$ and $\tau_{90}$(global) (see the left panel of Fig.~\ref{obstaufig}), parameterizing the overall SFH of the target galaxies, with $p$-values $\lesssim$0.001 regardless of assumed stellar evolutionary model.  Such a correlation is predicted by multiple sets of cosmological simulations \citep{graus19,riggs24} as a consequence of the interplay between feedback-induced outward radial reshuffling of stellar orbits (causing steeper gradients) and an increasing galactocentric radius where recent star formation occurs (flattening radial gradients) as discussed in Sect.~\ref{introsimsect}.  A detailed comparison against simulations is presented below in Sect.~\ref{simsect}.  

\item The only other parameter showing a moderately significant correlation ($p$-values$\leq$0.03 to within uncertainties) with $\gamma_{90}$ is the total stellar mass of the target galaxy, with similarly moderate correlation coefficients ($|$$\rho$$|$$\sim$0.35).  To illustrate this secondary correlation, our target galaxies in Fig.~\ref{obstaufig} are color-coded by stellar mass.  For clarity, not shown is one outlier well beyond the axis limits, LGS3, which is a transition dwarf (dTrans) that formed most of its mass very early on ($\tau_{90}$(global)=9.10$^{+1.84}_{-1.70}$ Gyr) with a steep outside-in age gradient ($\gamma_{90}$=8.33$^{+2.74}_{-2.47}$ Gyr/R$_{hl}$).  LGS3 is also the least massive galaxy in our sample (Log M$_{\star}$/M$_{\odot}$$=$5.96; \citealt{k18}) and we discuss age gradients at the low end of our sampled mass range in the context of simulation predictions in Sect.~\ref{simsect}.

\item For $\gamma_{50}$, we did not detect a statistically significant correlation with \textit{any} of the global galaxy properties examined, including $\tau_{50}$(global).  We demonstrate in Sect.~\ref{t50simsect} that this result can be used to discriminate between the predictions of different sets of simulations with different stellar feedback implementations.

\item Similarly, we find no significant correlation between $\gamma_{50}$ or $\gamma_{90}$ versus any of the environmental parameters $\theta_{1}$, D(NN), or D(NLG$_{\rm 10}$).  This indicates that environment does not play a major role in setting internal stellar age gradients in dwarfs over lookback times from $\sim$1$-$12 Gyr, in accord with simulations by \citet[][see Sect.~\ref{simsect}]{riggs24}.  The only known exceptions can be found in extreme cases where galaxies have been closely interacting for several Gyr such as the LMC and SMC \citep[e.g.,][and references therein]{cohen24a,cohen24b}.  

\end{itemize}

The full set of correlation coefficients for $\gamma_{90}$ and $\gamma_{50}$ versus global galaxy properties assuming SFH fits from PARSEC and MIST stellar evolutionary models are provided in Appendix~\ref{coefftabsect}.  

\begin{longrotatetable}
\setlength{\tabcolsep}{3pt}
\begin{deluxetable}{lllccccccccccccccccc}
\tabletypesize{\tiny} 
\tablecaption{Structural Parameters and Radial Age Gradient Slopes \label{slopetab}}
\tablehead{
\colhead{} & \multicolumn{7}{c}{Structural Parameters} & \multicolumn{6}{c}{Assuming PARSEC Models} & \multicolumn{6}{c}{Assuming MIST Models} \\
\colhead{Galaxy} & \colhead{RA} & \colhead{Dec.} & \colhead{R$_{hl}$} & \colhead{$h_{r}$} & \colhead{1-$\frac{b}{a}$} & \colhead{PA} & \colhead{Ref.} & \colhead{$\gamma_{90}$} & \colhead{$\gamma_{50}$} & \colhead{\shortstack{$\tau_{90}$\\(global)}} & \colhead{\shortstack{$\tau_{50}$\\(global)}} & \colhead{(m-M)$_{0,fit}$} & \colhead{$A_{V,fit}$} & \colhead{$\gamma_{90}$} & \colhead{$\gamma_{50}$} & \colhead{\shortstack{$\tau_{90}$\\(global)}} & \colhead{\shortstack{$\tau_{50}$\\(global)}} & \colhead{(m-M)$_{0,fit}$} & \colhead{$A_{V,fit}$} \\ \colhead{} & \colhead{hh:mm:ss} & \colhead{+dd:mm:ss} & \colhead{$\arcsec$} & \colhead{$\arcsec$} & \colhead{} & \colhead{$^{\circ}$} & \colhead{} & \colhead{Gyr/R$_{hl}$} & \colhead{Gyr/R$_{hl}$} & \colhead{Gyr} & \colhead{Gyr} & \colhead{mag} & \colhead{mag} & \colhead{Gyr/R$_{hl}$} & \colhead{Gyr/R$_{hl}$} & \colhead{Gyr} & \colhead{Gyr} & \colhead{mag} & \colhead{mag}}
\startdata
WLM & 00:01:59 & -15:27:27 & 278 & 162 & 0.56 & 183.6 & 1 & 1.43$^{+0.28}_{-0.29}$ & 2.04$^{+0.46}_{-0.47}$ & 1.74$^{+0.20}_{-0.17}$ & 6.00$^{+0.32}_{-0.32}$ & 24.93\tablenotemark{a} & 0.10\tablenotemark{a} & 1.78$^{+0.28}_{-0.28}$ & 2.10$^{+0.31}_{-0.34}$ & 1.84$^{+0.18}_{-0.16}$ & 6.60$^{+0.23}_{-0.19}$ & 24.93\tablenotemark{a} & 0.10\tablenotemark{a} \\
ESO410-005 & 00:15:31 & -32:10:48 & 25 & 18 & 0.37 & 57.0 & 3 & 1.14$^{+1.00}_{-0.80}$ & 1.85$^{+1.74}_{-1.98}$ & 2.15$^{+0.56}_{-0.42}$ & 8.19$^{+1.26}_{-1.04}$ & 26.48 & 0.00 & 1.11$^{+1.05}_{-0.98}$ & 2.33$^{+1.67}_{-1.66}$ & 1.91$^{+0.73}_{-0.46}$ & 7.64$^{+1.06}_{-1.23}$ & 26.53 & 0.00 \\
ESO294-010 & 00:26:33 & -41:51:19 & 30 & 15 & 0.51 & 8.0 & 3 & 2.05$^{+1.26}_{-1.25}$ & 1.20$^{+1.45}_{-1.56}$ & 2.42$^{+0.61}_{-0.51}$ & 10.55$^{+1.30}_{-1.25}$ & 26.49 & 0.00 & 2.33$^{+1.43}_{-1.40}$ & 0.25$^{+1.47}_{-1.57}$ & 2.33$^{+0.80}_{-0.50}$ & 10.68$^{+1.15}_{-0.98}$ & 26.59 & 0.00 \\
ESO540-032 & 00:50:24 & -19:54:24 & 36 & 21 & 0.60 & 325.0 & 1 & 0.96$^{+2.56}_{-2.56}$ & -0.52$^{+0.87}_{-0.86}$ & 5.29$^{+2.14}_{-1.72}$ & 12.73$^{+0.54}_{-0.53}$ & 27.75 & 0.01 & 1.44$^{+1.75}_{-1.70}$ & 1.15$^{+1.74}_{-1.66}$ & 3.84$^{+1.27}_{-1.03}$ & 8.91$^{+1.39}_{-1.02}$ & 27.85 & 0.01 \\
LGS3 & 01:03:55 & +21:53:06 & 126 & 75 & 0.20 & 0.0 & 3 & 8.33$^{+2.74}_{-2.47}$ & 2.41$^{+2.54}_{-2.53}$ & 9.10$^{+1.84}_{-1.70}$ & 10.35$^{+1.15}_{-1.02}$ & 24.13 & 0.01 & 8.26$^{+2.65}_{-2.37}$ & 3.53$^{+3.05}_{-2.70}$ & 9.40$^{+1.65}_{-1.56}$ & 11.88$^{+1.44}_{-1.08}$ & 24.03 & 0.06 \\
IC1613 & 01:04:47 & +02:07:04 & 288 & 228 & 0.15 & 80.0 & 5 & 0.64$^{+0.31}_{-0.31}$ & 0.83$^{+0.60}_{-0.56}$ & 1.86$^{+0.26}_{-0.18}$ & 6.45$^{+0.48}_{-0.38}$ & 24.39\tablenotemark{a} & 0.07\tablenotemark{a} & 0.49$^{+0.43}_{-0.48}$ & 0.44$^{+0.75}_{-0.78}$ & 2.09$^{+0.34}_{-0.31}$ & 7.19$^{+0.61}_{-0.55}$ & 24.39\tablenotemark{a} & 0.07\tablenotemark{a} \\
UGC685 & 01:07:22 & +16:41:05 & 33 & 20 & 0.36 & 116.9 & 1 & 0.61$^{+0.75}_{-0.66}$ & 1.13$^{+1.28}_{-1.37}$ & 1.13$^{+0.35}_{-0.20}$ & 2.85$^{+0.97}_{-0.56}$ & 28.31 & 0.21 & 1.64$^{+2.09}_{-1.89}$ & 3.17$^{+1.76}_{-2.01}$ & 2.89$^{+1.34}_{-0.73}$ & 6.41$^{+1.71}_{-0.90}$ & 28.31 & 0.06 \\
UGC1281 & 01:49:32 & +32:35:17 & 79 & 64 & 0.79 & 41.9 & 2 & -0.13$^{+0.55}_{-0.61}$ & 0.51$^{+1.35}_{-1.46}$ & 0.94$^{+0.29}_{-0.28}$ & 2.44$^{+0.79}_{-0.54}$ & 28.51 & 0.04 & 1.04$^{+2.01}_{-1.57}$ & 2.31$^{+2.37}_{-2.25}$ & 2.03$^{+1.11}_{-0.59}$ & 5.46$^{+1.79}_{-0.80}$ & 28.41 & 0.00 \\
Phoenix & 01:51:06 & -44:26:41 & 138 & 82 & 0.30 & 8.0 & 3,4 & 1.96$^{+0.46}_{-0.46}$ & 0.16$^{+0.79}_{-0.72}$ & 3.68$^{+0.30}_{-0.37}$ & 8.96$^{+0.63}_{-0.63}$ & 23.13\tablenotemark{a} & 0.04\tablenotemark{a} & 2.33$^{+0.54}_{-0.53}$ & -0.20$^{+0.71}_{-0.63}$ & 3.94$^{+0.37}_{-0.41}$ & 10.19$^{+0.57}_{-0.61}$ & 23.13\tablenotemark{a} & 0.04\tablenotemark{a}\\
NGC0784 & 02:01:17 & +28:50:07 & 95 & 57 & 0.72 & 0.9 & 1 & -0.23$^{+0.53}_{-0.66}$ & -0.05$^{+1.64}_{-2.06}$ & 0.79$^{+0.23}_{-0.24}$ & 2.45$^{+0.74}_{-0.75}$ & 28.58 & 0.11 & 0.75$^{+1.23}_{-1.01}$ & 0.47$^{+2.37}_{-2.09}$ & 1.31$^{+0.71}_{-0.35}$ & 4.23$^{+1.69}_{-0.67}$ & 28.58 & 0.06 \\
NGC2366 & 07:28:53 & +69:12:37 & 154 & 96 & 0.66 & 31.0 & 1 & -0.05$^{+1.08}_{-1.18}$ & -0.93$^{+4.18}_{-3.96}$ & 1.16$^{+0.73}_{-0.39}$ & 9.03$^{+1.69}_{-1.50}$ & 27.53 & 0.05 &  -0.52$^{+1.17}_{-1.55}$ & -0.20$^{+2.15}_{-2.48}$ & 1.31$^{+0.91}_{-0.49}$ & 4.22$^{+1.67}_{-0.82}$ & 27.58 & 0.00 \\
UGC4459 & 08:34:07 & +66:10:39 & 44 & 25 & 0.11 & 133.0 & 1 & 0.48$^{+1.53}_{-1.47}$ & 2.52$^{+2.30}_{-2.58}$ & 1.91$^{+1.34}_{-0.78}$ & 8.09$^{+1.88}_{-1.75}$ & 27.73 & 0.20 & 0.53$^{+1.67}_{-1.75}$ & 1.76$^{+2.24}_{-2.14}$ & 2.29$^{+1.34}_{-0.82}$ & 5.35$^{+1.72}_{-1.15}$ & 27.73 & 0.10 \\
UGC4483 & 08:37:03 & +69:46:32 & 25 & 14 & 0.40 & 351.5 & 1 & 0.83$^{+1.07}_{-0.98}$ & -0.83$^{+3.02}_{-2.67}$ & 1.17$^{+0.67}_{-0.35}$ & 9.01$^{+1.48}_{-1.65}$ & 27.67 & 0.00 &  0.57$^{+1.13}_{-1.30}$ & 1.59$^{+2.31}_{-2.68}$ & 1.28$^{+0.77}_{-0.42}$ & 5.85$^{+1.53}_{-1.16}$ & 27.72 & 0.00 \\
UGC5139 & 09:40:30 & +71:11:06 & 82 & 43 & 0.06 & 66.8 & 1 & 2.18$^{+3.19}_{-2.20}$ & 1.15$^{+2.44}_{-3.03}$ & 2.42$^{+1.96}_{-1.05}$ & 11.42$^{+0.93}_{-1.10}$ & 27.92 & 0.14 & 0.04$^{+2.27}_{-2.67}$ & 1.46$^{+3.07}_{-3.09}$ & 1.99$^{+1.30}_{-0.81}$ & 5.21$^{+1.95}_{-0.89}$ & 27.92 & 0.09 \\
UGC5364 & 09:59:25 & +30:44:49 & 100 & 52 & 0.52 & 102.5 & 1 & 0.94$^{+0.52}_{-0.60}$ & 1.52$^{+1.02}_{-1.13}$ & 1.24$^{+0.29}_{-0.22}$ & 3.93$^{+0.49}_{-0.45}$ & 24.45 & 0.01 & 0.98$^{+0.63}_{-0.78}$ & 1.90$^{+1.24}_{-1.14}$ & 1.23$^{+0.31}_{-0.33}$ & 4.21$^{+0.48}_{-0.47}$ & 24.40 & 0.01 \\
SextansB & 09:59:60 & +05:19:58 & 121 & 70 & 0.33 & 97.1 & 1 & -0.88$^{+1.99}_{-2.48}$ & 4.07$^{+5.10}_{-5.71}$ & 1.45$^{+0.70}_{-0.53}$ & 7.92$^{+1.84}_{-1.58}$ & 25.73 & 0.09 & -0.09$^{+4.50}_{-4.98}$ & -0.25$^{+5.36}_{-4.93}$ & 2.92$^{+1.62}_{-1.25}$ & 6.78$^{+1.76}_{-1.34}$ & 25.68 & 0.04 \\
NGC3109 & 10:03:10 & -26:09:29 & 242 & 144 & 0.72 & 94.3 & 1 & 0.25$^{+2.00}_{-1.83}$ & -1.92$^{+3.75}_{-3.36}$ & 1.79$^{+1.31}_{-0.59}$ & 6.89$^{+1.48}_{-1.81}$ & 25.63 & 0.13 & -0.63$^{+2.50}_{-2.75}$ & -3.39$^{+3.46}_{-2.85}$ & 2.56$^{+1.74}_{-0.95}$ & 8.29$^{+1.26}_{-1.72}$ & 25.58 & 0.08 \\
Antlia & 10:04:04 & -27:19:52 & 72 & 51 & 0.40 & 135.0 & 3 & 3.55$^{+1.76}_{-1.69}$ & 1.56$^{+2.30}_{-1.87}$ & 3.61$^{+1.09}_{-0.79}$ & 8.14$^{+1.08}_{-0.78}$ & 25.54 & 0.22 & 2.86$^{+1.90}_{-1.91}$ & 2.28$^{+2.21}_{-1.97}$ & 3.24$^{+1.17}_{-0.81}$ & 8.48$^{+1.18}_{-0.83}$ & 25.59 & 0.12 \\
SextansA & 10:11:01 & -04:41:38 & 113 & 51 & 0.14 & 46.3 & 1 & 0.78$^{+2.38}_{-2.10}$ & -5.02$^{+5.94}_{-4.81}$ & 1.17$^{+0.95}_{-0.78}$ & 7.72$^{+1.98}_{-2.35}$ & 25.84 & 0.07 & -0.15$^{+2.50}_{-2.73}$ & 4.98$^{+6.11}_{-6.15}$ & 1.03$^{+1.22}_{-0.85}$ & 6.85$^{+2.52}_{-2.49}$ & 25.69 & 0.07 \\
NGC3738 & 11:35:49 & +54:31:27 & 39 & 32 & 0.34 & 340.8 & 1 & 0.18$^{+0.39}_{-0.35}$ & 0.34$^{+0.87}_{-1.05}$ & 0.93$^{+0.41}_{-0.32}$ & 1.95$^{+1.19}_{-0.74}$ & 28.57 & 0.00 & 1.13$^{+1.15}_{-0.92}$ & 1.06$^{+1.70}_{-1.39}$ & 0.93$^{+0.74}_{-0.52}$ & 3.13$^{+1.41}_{-1.20}$ & 28.57 & 0.00 \\
NGC3741 & 11:36:06 & +45:17:11 & 23 & 14 & 0.32 & 12.3 & 1 & 1.95$^{+2.06}_{-1.58}$ & 1.51$^{+2.69}_{-2.78}$ & 1.81$^{+1.08}_{-0.65}$ & 9.37$^{+1.54}_{-1.64}$ & 27.49 & 0.02 & 0.51$^{+1.62}_{-1.64}$ & -0.60$^{+2.64}_{-2.18}$ & 1.91$^{+1.32}_{-0.81}$ & 7.50$^{+1.26}_{-1.20}$ & 27.49 & 0.02 \\
UGC6817 & 11:50:54 & +38:52:50 & 77 & 45 & 0.40 & 238.4 & 1 & 0.75$^{+1.92}_{-1.61}$ & 2.43$^{+4.74}_{-5.07}$ & 1.39$^{+0.85}_{-0.59}$ & 9.30$^{+1.59}_{-2.15}$ & 27.02 & 0.17 & 1.47$^{+3.71}_{-3.64}$ & 2.30$^{+3.92}_{-3.61}$ & 2.93$^{+2.31}_{-1.34}$ & 11.05$^{+1.10}_{-1.59}$ & 27.02 & 0.17 \\
NGC4068 & 12:04:02 & +52:35:27 & 57 & 33 & 0.44 & 31.4 & 1 & 1.35$^{+1.31}_{-0.91}$ & 4.88$^{+2.21}_{-2.59}$ & 1.22$^{+0.61}_{-0.45}$ & 5.03$^{+1.39}_{-1.18}$ & 28.16 & 0.11 & 0.45$^{+1.35}_{-1.46}$ & 0.09$^{+2.13}_{-2.20}$ & 1.77$^{+1.02}_{-0.64}$ & 4.25$^{+1.47}_{-0.95}$ & 28.11 & 0.11 \\
NGC4163 & 12:12:09 & +36:10:08 & 36 & 22 & 0.35 & 192.4 & 1 & 1.74$^{+1.76}_{-1.92}$ & 3.20$^{+1.82}_{-1.77}$ & 2.95$^{+1.24}_{-0.71}$ & 8.82$^{+1.71}_{-1.55}$ & 27.33 & 0.01 & 2.01$^{+1.86}_{-1.78}$ & 4.74$^{+1.95}_{-2.12}$ & 2.54$^{+1.29}_{-0.83}$ & 6.52$^{+1.50}_{-1.40}$ & 27.38 & 0.01 \\
NGC4190 & 12:13:44 & +36:38:01 & 54 & 34 & 0.09 & 34.1 & 2 & 0.55$^{+1.80}_{-1.25}$ & 6.69$^{+4.88}_{-4.91}$ & 1.03$^{+0.86}_{-0.46}$ & 4.28$^{+2.16}_{-2.09}$ & 27.82 & 0.18 & 4.81$^{+2.14}_{-2.12}$ & 7.67$^{+2.54}_{-2.56}$ & 4.03$^{+1.35}_{-1.09}$ & 11.66$^{+0.56}_{-0.58}$ & 27.62 & 0.03 \\
UGCA281 & 12:26:17 & +48:29:38 & 19 & 13 & 0.25 & 264.9 & 1 & 0.05$^{+0.53}_{-0.55}$ & 0.53$^{+1.49}_{-1.19}$ & 0.91$^{+0.49}_{-0.45}$ & 2.06$^{+0.92}_{-0.68}$ & 28.68 & 0.00 & -0.67$^{+1.13}_{-1.19}$ & -1.89$^{+1.18}_{-0.94}$ & 1.89$^{+1.15}_{-1.00}$ & 5.11$^{+0.94}_{-0.85}$ & 28.68 & 0.00 \\
UGC7577 & 12:27:42 & +43:29:33 & 100 & 58 & 0.50 & 127.5 & 1 & 1.26$^{+1.76}_{-1.30}$ & 7.20$^{+4.36}_{-4.63}$ & 1.44$^{+0.81}_{-0.57}$ & 9.29$^{+2.21}_{-2.04}$ & 26.98 & 0.01 & 1.79$^{+3.21}_{-2.77}$ & 4.84$^{+4.26}_{-4.57}$ & 2.71$^{+2.03}_{-1.19}$ & 7.02$^{+2.45}_{-1.80}$ & 26.98 & 0.01 \\
UGCA292 & 12:38:40 & +32:45:46 & 31 & 15 & 0.16 & 63.9 & 1 & 0.44$^{+1.32}_{-1.16}$ & 3.45$^{+2.51}_{-2.72}$ & 0.83$^{+0.80}_{-0.37}$ & 4.22$^{+1.47}_{-1.06}$ & 27.83 & 0.00 & 0.32$^{+1.53}_{-1.40}$ & 1.88$^{+2.32}_{-2.17}$ & 1.16$^{+1.04}_{-0.46}$ & 2.98$^{+1.41}_{-0.85}$ & 27.83 & 0.00 \\
UGC8024 & 12:54:05 & +27:08:55 & 49 & 28 & 0.46 & 215.9 & 1 & 0.66$^{+1.47}_{-1.09}$ & -3.91$^{+3.37}_{-2.82}$ & 1.06$^{+0.66}_{-0.48}$ & 7.25$^{+1.39}_{-1.83}$ & 28.03 & 0.00 & 0.14$^{+1.34}_{-1.19}$ & 0.48$^{+2.06}_{-1.84}$ & 0.96$^{+0.89}_{-0.46}$ & 3.00$^{+1.05}_{-0.88}$ & 28.03 & 0.00 \\
UGC8091 & 12:58:40 & +14:13:06 & 36 & 20 & 0.33 & 40.7 & 1 & 1.84$^{+1.57}_{-1.56}$ & 2.22$^{+1.82}_{-1.82}$ & 2.50$^{+0.76}_{-0.54}$ & 8.68$^{+1.11}_{-0.89}$ & 26.65 & 0.02 & 1.63$^{+1.79}_{-1.96}$ & 3.30$^{+2.05}_{-1.93}$ & 2.46$^{+0.89}_{-0.70}$ & 8.01$^{+1.21}_{-0.85}$ & 26.65 & 0.02 \\
UGCA320 & 13:03:17 & -17:25:25 & 69 & 42 & 0.83 & 115.0 & 2 & -0.07$^{+0.44}_{-0.51}$ & 0.11$^{+1.20}_{-1.37}$ & 0.80$^{+0.33}_{-0.30}$ & 2.13$^{+1.01}_{-0.65}$ & 28.80 & 0.12 & 0.61$^{+0.77}_{-0.60}$ & 1.16$^{+2.28}_{-1.88}$ & 0.71$^{+0.45}_{-0.27}$ & 3.39$^{+1.24}_{-1.01}$ & 28.80 & 0.12 \\
UGC8201 & 13:06:25 & +67:42:25 & 83 & 43 & 0.50 & 269.3 & 1 & 0.73$^{+1.42}_{-1.36}$ & -3.19$^{+2.61}_{-2.49}$ & 1.24$^{+0.82}_{-0.52}$ & 8.99$^{+0.78}_{-1.79}$ & 28.37 & 0.02 & 1.22$^{+1.72}_{-1.59}$ & 3.06$^{+2.63}_{-2.71}$ & 1.50$^{+1.09}_{-0.55}$ & 3.98$^{+1.26}_{-0.92}$ & 28.32 & 0.02 \\
UGC8508 & 13:30:44 & +54:54:38 & 36 & 22 & 0.37 & 118.1 & 1 & 1.39$^{+1.47}_{-1.49}$ & -1.25$^{+2.31}_{-1.95}$ & 1.90$^{+0.64}_{-0.58}$ & 10.55$^{+0.85}_{-1.55}$ & 27.13 & 0.00 & 0.94$^{+1.68}_{-1.83}$ & 2.36$^{+2.54}_{-2.55}$ & 2.10$^{+0.94}_{-0.69}$ & 6.25$^{+1.36}_{-1.18}$ & 27.13 & 0.00 \\
UGC8638 & 13:39:19 & +24:46:34 & 38 & 22 & 0.45 & 69.2 & 1 & 1.15$^{+1.14}_{-0.82}$ & 0.40$^{+2.09}_{-1.86}$ & 1.50$^{+0.57}_{-0.38}$ & 4.98$^{+1.44}_{-1.26}$ & 28.11 & 0.00 & 0.95$^{+1.98}_{-1.89}$ & 3.06$^{+2.16}_{-2.51}$ & 3.00$^{+1.48}_{-0.80}$ & 6.10$^{+1.67}_{-1.00}$ & 28.06 & 0.00 \\
UGC8651 & 13:39:54 & +40:44:25 & 52 & 29 & 0.47 & 241.1 & 1 & -0.13$^{+1.63}_{-1.79}$ & -1.10$^{+3.37}_{-2.93}$ & 2.06$^{+1.20}_{-0.64}$ & 8.84$^{+1.54}_{-1.45}$ & 27.41 & 0.00 & 0.95$^{+1.98}_{-1.84}$ & 1.69$^{+2.38}_{-2.10}$ & 2.15$^{+1.09}_{-0.70}$ & 4.84$^{+1.43}_{-0.84}$ & 27.41 & 0.00 \\
UGC8837 & 13:54:46 & +53:54:12 & 64 & 49 & 0.69 & 19.7 & 2 & 0.09$^{+0.46}_{-0.42}$ & 0.08$^{+1.11}_{-1.53}$ & 1.02$^{+0.35}_{-0.19}$ & 2.40$^{+0.98}_{-0.65}$ & 29.15 & 0.00 & 0.77$^{+1.28}_{-1.14}$ & 0.39$^{+1.91}_{-1.80}$ & 1.47$^{+0.84}_{-0.51}$ & 4.30$^{+1.64}_{-0.82}$ & 29.10 & 0.00 \\
UGC9128 & 14:15:57 & +23:03:23 & 36 & 21 & 0.40 & 224.7 & 1 & 1.00$^{+0.94}_{-0.91}$ & 0.98$^{+2.55}_{-2.63}$ & 1.06$^{+0.56}_{-0.36}$ & 5.45$^{+1.76}_{-1.24}$ & 26.71 & 0.11 & 0.95$^{+0.83}_{-0.80}$ & 1.26$^{+1.87}_{-2.12}$ & 1.25$^{+0.61}_{-0.30}$ & 4.39$^{+1.45}_{-0.79}$ & 26.71 & 0.01 \\
UGC9240 & 14:24:44 & +44:31:36 & 43 & 24 & 0.20 & 127.7 & 1 & 0.78$^{+1.52}_{-1.45}$ & -0.08$^{+2.64}_{-2.28}$ & 1.85$^{+1.15}_{-0.57}$ & 9.56$^{+1.28}_{-1.51}$ & 27.21 & 0.00 & 0.35$^{+1.39}_{-1.76}$ & 3.08$^{+2.27}_{-2.35}$ & 2.04$^{+1.18}_{-0.76}$ & 5.66$^{+1.52}_{-0.91}$ & 27.26 & 0.00 \\
IC4662 & 17:47:08 & -64:38:31 & 34 & 21 & 0.35 & 280.9 & 1 & 1.13$^{+1.12}_{-0.78}$ & 5.64$^{+1.33}_{-1.59}$ & 0.87$^{+0.54}_{-0.33}$ & 3.55$^{+1.64}_{-1.19}$ & 26.93 & 0.29 & 1.15$^{+1.62}_{-1.48}$ & 1.45$^{+2.38}_{-2.20}$ & 1.91$^{+1.38}_{-0.72}$ & 7.57$^{+2.05}_{-1.74}$ & 26.93 & 0.24 \\
NGC6822 & 19:44:59 & -14:47:35 & 612 & 364 & 0.14 & 75.8 & 6 & 0.70$^{+0.28}_{-0.30}$ & 1.02$^{+0.60}_{-0.57}$ & 2.57$^{+0.19}_{-0.15}$ &  7.14$^{+0.55}_{-0.48}$ & 23.54\tablenotemark{a} & 0.54\tablenotemark{a} & 0.96$^{+0.29}_{-0.31}$ & 0.67$^{+0.58}_{-0.55}$ & 2.40$^{+0.25}_{-0.20}$ & 7.82$^{+0.53}_{-0.53}$ 23.54\tablenotemark{a} & 0.54\tablenotemark{a} \\
DDO210 & 20:46:52 & -12:50:53 & 88 & 63 & 0.50 & 99.0 & 3 & 2.85$^{+1.01}_{-0.94}$ & 1.73$^{+1.73}_{-1.54}$ & 3.33$^{+0.45}_{-0.43}$ & 7.40$^{+0.95}_{-0.71}$ & 24.90 & 0.04 & 2.94$^{+1.19}_{-1.09}$ & 1.66$^{+1.86}_{-1.58}$ & 3.48$^{+0.57}_{-0.52}$ & 7.93$^{+0.98}_{-0.83}$ & 24.85 & 0.04 \\
IC5152 & 22:02:42 & -51:17:47 & 64 & 38 & 0.41 & 279.3 & 1 & 0.35$^{+0.56}_{-0.51}$ & 3.49$^{+2.16}_{-2.19}$ & 0.77$^{+0.38}_{-0.33}$ & 3.57$^{+1.44}_{-0.80}$ & 26.36 & 0.07 & 1.11$^{+0.96}_{-0.74}$ & -2.38$^{+3.05}_{-2.60}$ & 0.89$^{+0.47}_{-0.35}$ & 6.95$^{+2.15}_{-2.29}$ & 26.36 & 0.07 \\
\enddata
\tablenotetext{a}{For galaxies with multiple non-contiguous pointings, $(m-M)_{0}$ and $A_{V}$ were held fixed for SFH fitting; this choice did not affect the results beyond their uncertainties.}
\tablecomments{References: (1) McQuinn et al.~(2025, in prep.); (2) \citet{lelli16}; (3) \citet{m12}; (4) \citet{battaglia12}; (5) \citet{bernard07}; (6) \citet{tantalo22} (7) \citet{kirby08}}
\end{deluxetable}
\end{longrotatetable}

\begin{figure}
\gridline{\fig{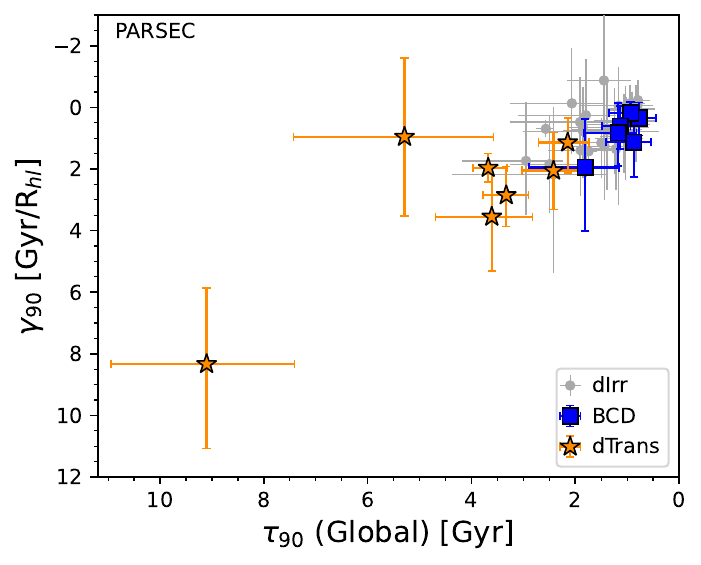}{0.49\textwidth}{}
          \fig{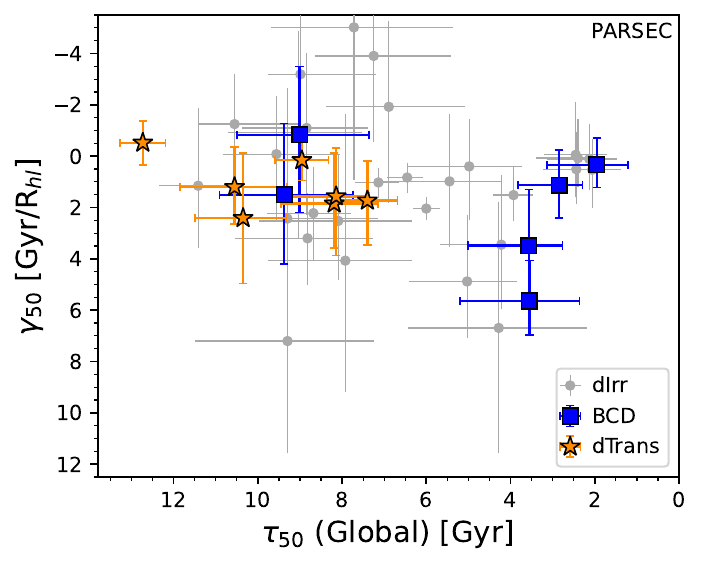}{0.49\textwidth}{}}
\caption{As in Fig.~\ref{obstaufig}, but with galaxies color-coded by morphological type.  The steeper values of $\gamma_{90}$ for the transition dwarfs in our sample is consistent with their earlier values of $\tau_{90}$(global) and lower stellar masses, while BCDs do not seem to have preferentially steep stellar age gradients (see text for discussion).   
\label{morphtypefig}}
\end{figure}

Our sample also provides the opportunity to assess any relationship between radial gradient slopes and morphological type.  While our goal of sampling the Local Volume (rather than solely Local Group) environment results in a sample comprised predominantly (67\%) of dIrrs (i.e., due to the morphology-density relation; \citealt{geha06,weisz11}), in Fig.~\ref{morphtypefig} we reproduce Fig.~\ref{obstaufig}, but with targets color-coded by morphology to highlight the loci of dTrans and BCDs.  These morphological types separate more clearly with respect to $\tau_{90}$(global) rather than $\tau_{50}$(global), consistent with the idea that morphology observed at the present day may be largely set in the most recent Gyr or so \citep{weisz11}.  Accordingly, the dTrans, which form larger fractions of their mass at earlier times, tend to have the steepest values of $\gamma_{90}$.  However, the dTrans with relatively more recent values of $\tau_{90}$(global) do adhere to $\gamma_{90}$-$\tau_{90}$(global) correlation we report above, with flatter values of $\gamma_{90}$, confirming that such a trend extends down to Log M$_{\star}$/M$_{\odot}$$\sim$6.  

Unlike the dTrans, all of the BCDs in our sample have fairly flat stellar age gradients (0$<$$\gamma_{90}$$<$2), consistent with their $\tau_{90}$(global)$\lesssim$2 Gyr.  This supports the idea that BCDs have fairly constant lifetime SFHs similar to dIrrs 
\citep[e.g.,][]{sacchi21}, in contrast to previous suggestions of steep outside-in gradients \citep[e.g.,][]{skillman03}.  Interestingly, a strong central concentration of mass has been inferred for multiple BCDs based on steeply rising rotation curves \citep{lelli12a,lelli12b}.  At face value, this might suggest steep stellar age gradients since simulations predict that the potential fluctuations driving steep outside-in age gradients also affect the dark matter distribution, transforming dark matter profiles from cuspy to cored for dwarfs with Log M$_{\star}$/M$_{\odot}$$\gtrsim$6 \citep{riggs24}.  However, mass modeling in BCDs by \citet{lelli12a,lelli12b} suggests that baryons may constitute as much as half of the total mass in their inner regions, and further modeling work is needed to compare this finding with the latest simulations.   

\subsection{Comparison with Age Gradients from the Literature \label{litsect}}

A primary motivation for this study is the lack of self-consistent, quantitative radial age gradient measurements for Local Volume dwarfs.  Among the few existing studies, \citet{hidalgo13} measured gradients in $\tau_{10}$ and $\tau_{95}$ for four relatively isolated late-type dwarfs near the edge of the Local Group  using an approach schematically similar to ours (i.e., fitting SFHs to CMDs in spatially selected radial bins).  There are two targets in common between our sample and \citet{hidalgo13}, Phoenix and LGS3.  To compare our results with theirs and place them in context, we calculate $\gamma_{95}$, which is the radial gradient in $\tau_{95}$ normalized to R$_{hl}$.  We recover the \citet{hidalgo13} result that, using $\tau_{95}$ to measure age gradients, Phoenix and LGS3 have quite steep slopes compared to other dwarfs, with values of $\gamma_{95}$ steeper than 92\% of our sample.  Using the distances and half-light radii from \citet{hidalgo13} in an attempt to directly compare measurements of $\delta$$\tau_{95}$/$\delta$d$_{\rm SMA}$ in terms of projected angular distance, the gradients we find are somewhat shallower than theirs ($\gamma_{95}$=2.9$^{+1.0}_{-1.0}$ Gyr/arcmin for LGS3 and $\gamma_{95}$=0.8$^{+0.2}_{-0.2}$ Gyr/arcmin for Phoenix).  However, this is likely due in large part to differences in methodology, including assumed structural parameters and different ranges of galactocentric distance within each galaxy used to designate radial bins in the \citet{hidalgo13} study.  

Quantitative constraints on radial age stellar age gradients were also obtained on a case-by-case basis for four other Local Group dwarfs.  First, for WLM, \citet{cohen25} combined the HST fields we analyze here with JWST NIRCam and NIRISS fields, finding a value of $\gamma_{50}$ consistent with ours to within uncertainties and a slightly flatter value of $\gamma_{90}$.  This discrepancy is likely due to a combination of different spatial sampling and smaller reported uncertainties in the \citet{cohen25} study resulting from the use of deep JWST imaging, for which a prescription accounting for systematic uncertainties in SFH fits \citep[cf.][]{dolphin_syserr} cannot currently be applied (which is why archival JWST fields have not been included here).

Second, radial stellar population gradients in the blue compact dwarf UGC 4483 were discussed by \citet{sacchi21}, finding $\delta$$\tau_{50}$/$\delta$R$_{SMA}$$\approx$3 Gyr/kpc and $\delta$$\tau_{90}$/$\delta$R$_{SMA}$$\approx$1.5 Gyr/kpc, or $\sim$0.6 Gyr/R$_{hl}$ and $\sim$1.3 Gyr/R$_{hl}$ respectively, consistent with our results to within uncertainties (see Table \ref{slopetab}).   
Third, for NGC 6822, \citet{cannon12} and \citet{fusco14} examined radial trends in $\tau_{75}$ and $\tau_{95}$ using six fields, finding mostly flat radial gradients.  Analyzing nearly triple the number of HST pointings, we find gradients that are outside-in but shallow (see Table \ref{slopetab}).  We also confirm the \citet{fusco14} result that there are individual fields that deviate from the overall radial trend in $\tau_{90}$.  \citet{fusco14} hypothesize that these deviations may be evidence of nascent spiral structure, supported by both their tendency to cluster in position angle-galactocentric radius space and, comparing trends in $\tau_{90}$ versus $\tau_{50}$, the increased amplitude of the deviations at more recent lookback times (see the bottom row of Figs. \ref{indslopes_parsec_t90fig}$-$\ref{indslopes_mist_t50fig}).  Lastly, we note that present-day outside-in radial gradients in $\tau_{90}$ and $\tau_{50}$ were measured in the SMC, although a direct comparison to our sample is complicated by its dramatic, prolonged interaction history with the LMC and resultant highly irregular structure \citep[][and references therein]{cohen24b}.    

\subsection{Comparison with Simulations \label{simsect}}

The simulations of both \citet{graus19} and \citet{riggs24} make quantitative predictions for dwarf galaxy radial age gradient slopes $\gamma_{90}$ and $\gamma_{50}$, and both studies analyze correlations (or the lack thereof) versus global mass assembly history parameterized by $\tau_{90}$(global) and $\tau_{50}$(global).  These predicted correlations are compared with our observations in Figs.~\ref{compsim90fig}-\ref{compsim50fig}, which we discuss in turn below, highlighting agreements and disagreements with simulation predictions and their implications.  

\subsubsection{$\gamma_{90}$ Versus $\tau_{90}$(global) \label{t90sect}}

\begin{figure}
\gridline{\fig{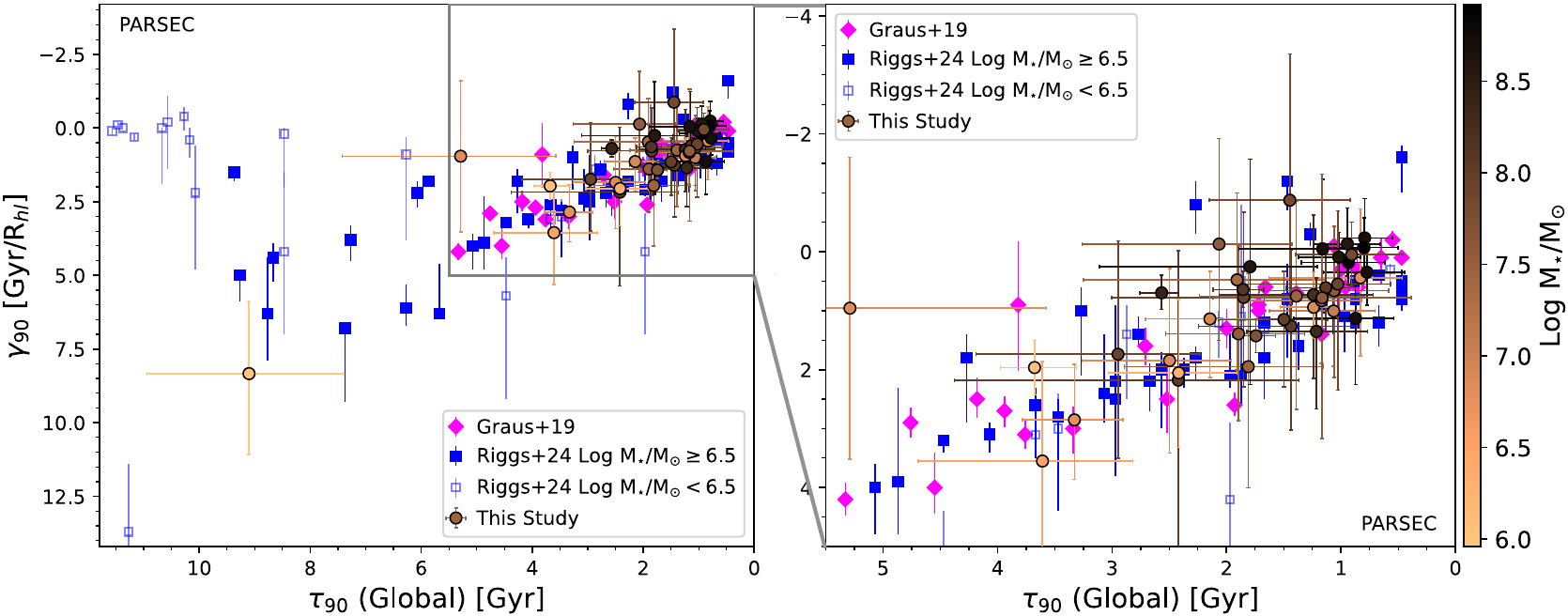}{0.99\textwidth}{}}
\caption{As in Fig.~\ref{obstaufig}, but with simulation predictions from \citet{graus19} and \citet{riggs24} overplotted.  For clarity, the right-hand panel shows a magnified portion of the left-hand panel.  The model predictions are in good agreement with each other and with the observations over the sampled mass range (see Sect.~\ref{t90sect}).
\label{compsim90fig}}
\end{figure}

In Fig.~\ref{compsim90fig}, we compare our observed $\gamma_{90}$ values as a function of $\tau_{90}$(global) as in Fig.~\ref{obstaufig}, now with simulation predictions from \citet{graus19} and \citet{riggs24} overplotted.  The \citet{riggs24} simulations, by design, include galaxies extending to lower stellar masses (Log M$_{\star}$/M$_{\odot}$=5.26) than either our observations (5.96$\leq$Log M$_{\star}$/M$_{\odot}$$\leq$8.92) or the \citet{graus19} simulations (5.67$\leq$Log M$_{\star}$/M$_{\odot}$$\leq$8.73), so we analyze a subset of the \citet{riggs24} dwarfs with Log M$_{\star}$/M$_{\odot}$$\geq$6.5 (shown using filled squares in Fig.~\ref{compsim90fig}).  We impose this division in the \citet{riggs24} sample because their simulated galaxies with Log M$_{\star}$/M$_{\odot}$$<$6.5 (shown as open squares in Fig.~\ref{compsim90fig}) tend to quench at early times, and such galaxies are poorly represented in our observational sample: dwarfs with Log M$_{\star}$/M$_{\odot}$$<$6.5 represent 35\% of the \citet{riggs24} sample but only 7\% of our observational sample (and 12\% of the \citealt{graus19} sample).  However, for comparison, we also provide results using the full \citet{riggs24} sample below (e.g., Fig.~\ref{violinfig} and Appendix~\ref{coefftabsect}).  

Fig.~\ref{compsim90fig} illustrates that the \citet{graus19} and \citet{riggs24} simulations predict correlations between $\gamma_{90}$ and $\tau_{90}$(global) that are in good agreement with each other as well as with the observations - there are \textit{no} galaxies in our observed sample with age gradient slopes falling clearly outside the loci defined by the simulations in Fig.~\ref{compsim90fig} given observational uncertainties.  The only target in our sample with a quenching time more than $\sim$6 Gyr ago (LGS3, $\tau_{90}$(global)$\sim$9 Gyr) falls outside the range of $\tau_{90}$(global) sampled by the \citet{graus19} simulations, but its steep outside-in radial age gradient is consistent with the predictions of the \citet{riggs24} simulations in the relevant stellar mass range.  While \citet{riggs24} find that many of their lowest-mass galaxies quench early and have flat age gradients, they also report one simulated galaxy (seen in the lower left corner of Fig.~\ref{compsim90fig}) with an even steeper radial gradient ($\gamma_{90}$=13.8 Gyr/R$_{hl}$).  They find that the steep gradient results from a predominance of old stars that have migrated to large radii, complemented by a small fraction ($\lesssim$15\%) of young ($<$1 Gyr) stars close to the center that are completely lacking beyond 2R$_{hl}$, and this simulated galaxy is nearly identical to LGS3 in present-day total stellar mass (Log M$_{\star}$/M$_{\odot}$$=$5.95).  

To quantitatively compare observed versus simulation-predicted  correlations between $\gamma_{90}$ and $\tau_{90}$(global), we calculate the simulation-predicted Pearson $\rho$ and the Spearman $r_{s}$ (and $p$-values) as in Sect.~\ref{obscorrsect}, and compare them with the observed values in Fig.~\ref{violinfig}.  While the observations show somewhat smaller correlation coefficients than the simulations (we return to this point in Sect.~\ref{caveatsect}), our results assuming either set of stellar evolutionary models and both sets of simulations consistently predict a positive linear correlation between $\gamma_{90}$ and $\tau_{90}$(global) that is highly statistically significant ($p$-values $\lesssim$10$^{-3}$).  Meanwhile, the comparison of coefficients obtained using our mass-selected subsample of the \citet{riggs24} dwarfs (shown in blue in Fig.~\ref{violinfig}) versus the full \citet{riggs24} sample (shown in cyan in Fig.~\ref{violinfig}) illustrates the need to exclude the lowest-mass simulated dwarfs when comparing to the \citet{riggs24} simulations, since many of these galaxies quench at early times in their simulations, generating a \enquote{U} shape in the left panel of Fig.~\ref{compsim90fig} that results in low absolute values of the correlation coefficients (see \citealt{riggs24} for a more detailed discussion of age gradients in dwarfs with Log M$_{\star}$/M$_{\odot}$$<$6).

\subsubsection{$\gamma_{50}$ Versus $\tau_{50}$(global): Discriminating Between Simulations \label{t50simsect}}

\begin{figure}
\gridline{\fig{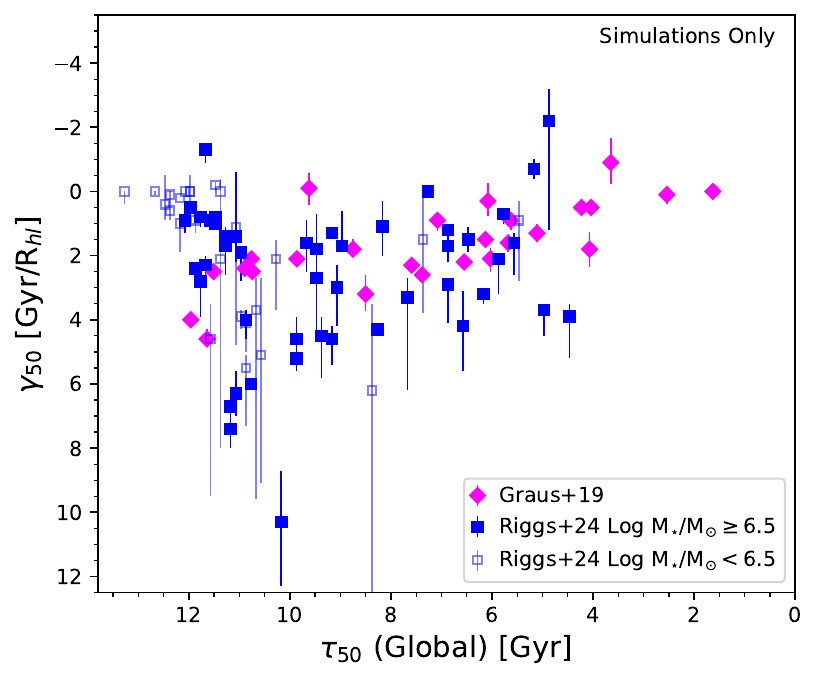}{0.47\textwidth}{}
          \fig{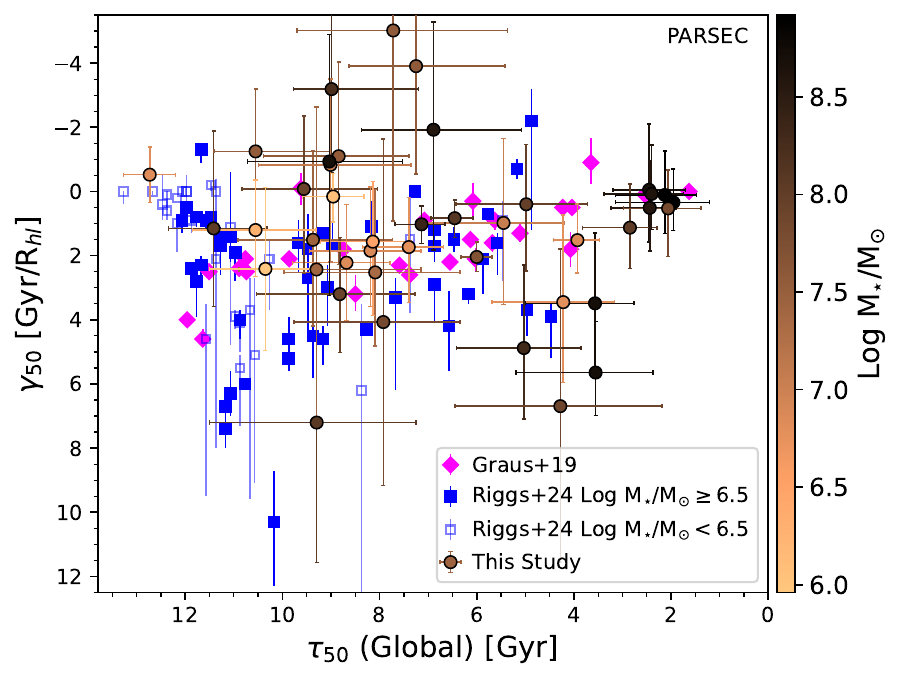}{0.51\textwidth}{}}
\caption{\textbf{Left:} Simulation predictions for radial gradient slope $\gamma_{50}$ as a function of $\tau_{50}$ (global) for simulated galaxies from \citet{graus19} and \citet{riggs24}, illustrating the weaker correlation between these two quantities predicted by the \citet{riggs24} simulations.  Symbols are as in Fig.~\ref{compsim90fig}.  \textbf{Right:} As in the left panel, but with data overplotted, color coded by stellar mass as in Figs.~\ref{obstaufig}-\ref{compsim90fig}.  
\label{compsim50fig}}
\end{figure}

We compare our observed values for $\gamma_{50}$ versus $\tau_{50}$(global) in Fig.~\ref{compsim50fig}.  For clarity, in the left-hand panel of Fig.~\ref{compsim50fig} we plot only the model predictions from \citet{graus19} and \citet{riggs24}.  Unlike in the case of $\gamma_{90}$, the two simulations clearly differ in the strength of the predicted correlations between $\gamma_{50}$ versus $\tau_{50}$(global), with the \citet{graus19} simulations predicting a correlation similar to that seen for $\gamma_{90}$ ($\rho$, $r_{s}$$\approx$0.7, $p$-values$<$10$^{-4}$) while the \citet{riggs24} simulations predict no significant correlation between $\gamma_{50}$ and $\tau_{50}$(global).  \citet{riggs24} interpret this difference as indicating that age gradients in the \citet{graus19} simulations are set much earlier, prior to $\tau_{50}$(global), explaining their coherent behavior at lookback times as early as 8$-$12 Gyr ago.  This behavior is traced by \citet{riggs24} to a difference in the implementation of stellar feedback.  Specifically, the \enquote{breathing modes} consisting of radial inflows and outflows of star-forming gas seen in the FIRE-2 simulations examined by \citet{graus19} can boost the outward radial reshuffling of stellar orbits, setting radial age gradients at earlier times.  Conversely, the \citet{riggs24} simulations, lacking radial velocity fluctuations among forming stars, find that the relationship between internal age gradients and global mass assembly history is established later, after $\tau_{50}$(global) but before $\tau_{90}$(global).  

\begin{figure}
\gridline{\fig{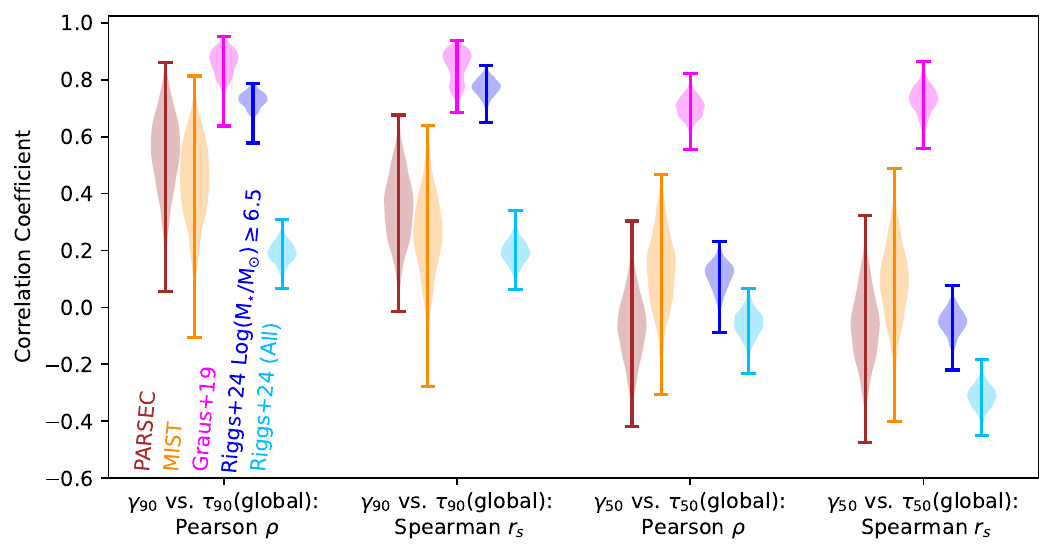}{0.99\textwidth}{}}
    \caption{A comparison of Pearson linear correlation coefficients and Spearman rank correlation coefficients measured from our observations (assuming PARSEC or MIST stellar evolutionary models shown in brown and orange respectively) with values measured from cosmological simulations by \citet{graus19} and \citet{riggs24} (shown in magenta and blue respectively).  In cyan, we also show results using the full \citet{riggs24} sample without removing quenched low-mass (Log M$_{\star}$/M$_{\odot}$$<$6.5) galaxies poorly represented in our observational sample.  While there is some tension in the comparison between observations and simulations for the correlation between $\gamma_{90}$ and $\tau_{90}$(global) (see Sect.~\ref{caveatsect}), the lack of observed correlation between $\gamma_{50}$ and $\tau_{50}$(global) indicates a clear preference for the \citet{riggs24} simulations over those of \citet{graus19}. \label{violinfig}}
\end{figure}

The observed values of $\gamma_{50}$ versus $\tau_{50}$(global) from Fig.~\ref{obstaufig} are compared to both sets of simulation predictions in the right-hand panel of Fig.~\ref{compsim50fig}.  Despite observational uncertainties, $\gamma_{50}$ and $\tau_{50}$(global) appear less correlated than predicted by the \citet{graus19} simulations, similar to the \citet{riggs24} predictions.  To quantify this result, the observed correlation coefficients (assuming both sets of stellar evolutionary models) are compared to both sets of model predictions in Fig.~\ref{violinfig}.  The uncertainties of the correlation coefficients shown via the distributions in Fig.~\ref{violinfig} (and also the $p$-values, both provided in Tables \ref{coefftab90} and \ref{coefftab50} in the Appendix) were obtained via a monte carlo procedure, in which the values of $\tau_{50}$, $\tau_{90}$, $\gamma_{50}$ and $\gamma_{90}$ for each galaxy were perturbed by draws from an (asymmetric) Gaussian with standard deviations equal to their 16th and 84th percentile uncertainties.  We find a complete lack of overlap between the observed $\gamma_{50}$-$\tau_{50}$(global) distributions versus the \citet{graus19} simulations.  Given that these distributions are based on 10,000 monte carlo iterations, this complete lack of overlap indicates that, while large observational uncertainties allow for values in Fig.~\ref{compsim50fig} that are not inconsistent with simulations, the probability of a $\gamma_{50}$-$\tau_{50}$(global) correlation as strong as predicted by \citet{graus19} \textit{across our entire sample of target galaxies} is $<$0.01\%, with a clear preference for the \citet{riggs24} simulations.  
As an additional check on this result, we have reperformed our analysis selecting the half of our target galaxy sample with the lowest uncertainties on $\gamma_{50}$, corresponding to an average (i.e., arithmetic mean of the positive and negative uncertainties) $\sigma$($\gamma_{50}$)$<$2.0 (2.16) Gyr/R$_{hl}$ assuming PARSEC (MIST) evolutionary models, finding that our data still exclude the \citet{graus19} predictions at the 99.98\% level.  
Importantly, such a comparison illustrates that, given the requisite quantity and quality of observational data, radial stellar age gradients are an actionable parameter to discriminate between different simulations with different stellar feedback implementation.  In particular, our results argue against the \enquote{breathing modes} that drive radial velocity fluctuations in gas hosting recent star formation.

\subsubsection{Other Global Galaxy Properties}

Beyond correlations with $\tau_{90}$(global) and $\tau_{50}$(global), which parameterize the global lifetime SFHs of our target galaxies, simulations make additional predictions regarding the relationships between global galaxy properties and radial stellar age gradient slopes that we may now confront.  In particular, there are two broad simulation predictions that agree with our results, which we highlight here as they also serve as observational constraints to be met by future modeling work.

First, the lack of correlation between radial gradient slopes versus any of the environmental metrics we explored (see Sect.~\ref{obscorrsect} and Appendix~\ref{coefftabsect}) is consistent with the predictions of \citet{riggs24}.  They found no clear difference between radial age gradient slopes in field dwarfs ($\gtrsim$1.5 Mpc from a Milky Way-mass host) versus those in group environments (within $\sim$1 Mpc of a Milky Way-mass disk galaxy), reinforcing the idea that radial age gradients are driven by internal, rather than external processes.  While our observational uncertainties are not insignificant, particularly for $\tau_{50}$(global) and $\gamma_{50}$, the uncertainties of $p$-values resulting from our monte carlo procedure (which explicitly accounts for asymmetric uncertainties on both $\tau_{50}$(global) and $\gamma_{50}$) exclude statistically significant correlations at the level of 1-$\sigma$ or more (see Table \ref{coefftab50}), implying that the lack of correlation is not due to large uncertainties alone.  Furthermore, examining the subset of simulated dwarfs that underwent mergers, \citet{riggs24} predict that even major mergers do not correlate with present-day stellar age gradients since the radial distribution of accreted stars and gas varies widely from galaxy to galaxy (also see \citealt{graus19}).  This is at odds with the predictions of \citet{benitezllambay16}, in which dwarfs form outside-in, with the older, more external component being accreted from merger events (also see \citealt{mostoghiu18}), although their simulated galaxies cover a relatively small range of present-day stellar mass (6.8$\leq$Log M$_{\star}$/M$_{\odot}$$\leq$7.8).  Within our sample, there is one galaxy (UGCA320) that may have interacted with a nearby companion (UGCA319, separated by $\sim$33 kpc; \citealt{karachentsev17} and references therein), and its values of $\gamma_{90}$ and $\gamma_{50}$ are typical of other targets with its global SFH (see Table \ref{slopetab}).  More detailed studies of the spatially resolved SFHs of interacting dwarfs may reveal the circumstances under which dwarf-dwarf interactions begin to affect internal stellar age gradients short of extreme cases like the LMC and SMC.

A second broad observational feature consistent with both the \citet{graus19} and \citet{riggs24} simulations is that 
\textit{all} of the galaxies in our sample show age gradient slopes ranging from flat to outside-in.  Although there exist galaxies with best-fit slopes that are formally \enquote{inside-out} ($\gamma_{50}$$<$0), none of these exceed flat gradients ($\gamma_{50}$=0) at a statistically significant ($\gtrsim$1.1$\sigma$) level.  Both sets of simulations find that flat age gradients can be produced via recent star formation out to a maximum radius that increases with time up to the present day.  Our observations strongly support predictions that this phenomenon may counterbalance, but does not outpace, outward radial migration of older stars over time due to feedback-induced potential well fluctuations.

\subsection{Caveats \label{caveatsect}}

We have kept our analysis as self-consistent as possible for the dual purposes of both assessing correlations within our sample (Sect.~\ref{obscorrsect}) as well as comparing our results to predictions from multiple sets of cosmological zoom-in simulations (Sect.~\ref{simsect}).  However, there do exist minor differences in the methodology used to calculate age gradients across the two sets of simulations and our observations.  We have thus far assumed these differences to have a negligible impact on our results, but we describe them here as they may be relevant to future analyses of additional cosmological simulations and/or additional observational targets.   

Comparing the \citet{riggs24} and \citet{graus19} simulations, there are three salient differences in the way that age gradients were assessed: First, \citet{riggs24} excluded halos smaller than 0.25 kpc to ensure adequate resolution, although this is unlikely to impact the stellar mass range of the dwarfs we examine given their size-luminosity relation \citep[e.g.,][]{applebaum21}.  Second, \citet{graus19} calculate their $\tau_{90}$ and $\tau_{50}$ values relative to the stellar mass of each galaxy at the present day (i.e., $z$=0), whereas \citet{riggs24} calculate cumulative mass fractions over the lifetime of the entire galaxy (i.e., including stars whose evolution has terminated before the present day), which is the approach we have taken observationally via the use of CSFHs (Sect.~\ref{sfhsect}).  Third, \citet{graus19} calculated their $\tau_{90}$(global) and $\tau_{50}$(global) values excluding particles outside 10\% of the virial radius of their simulated dwarfs.  This is in contrast to \citet{riggs24}, who used the population of the entire galaxy and fit radial gradients out to 1.5R$_{hl}$, whereas we have adopted a fixed cut in galactocentric radius of d$_{\rm SMA}$$\leq$4.4$h_{r}$ for practical reasons (see Sect.~\ref{structuralsect}).  This cut corresponds to $\sim$2.6R$_{hl}$ assuming an exponential profile, although using the four targets with relatively well-constrained age gradient slopes from multiple pointings (see Sect.~\ref{speccasesect}) we found that restricting our fits to d$_{\rm SMA}$$\leq$1.5R$_{hl}$ did not affect $\gamma_{90}$ or $\gamma_{50}$ beyond their uncertainties.    

Our comparison between age gradient slopes and simulation predictions may also be impacted to some extent by differences in the parameter space covered by the simulated galaxies compared to our observational sample with respect to both present-day stellar mass and global mass assembly history.  Our observational sample tends to contain more massive galaxies, with a median Log M$_{\star}$/M$_{\odot}$=7.83 and standard deviation $\sigma$=0.79 compared to the \citet{graus19} sample (median Log M$_{\star}$/M$_{\odot}$=7.16, $\sigma$=0.82) or the \citet{riggs24} sample, which has a median Log M$_{\star}$/M$_{\odot}$=7.37 ($\sigma$=0.71) even after requiring Log M$_{\star}$/M$_{\odot}$$>$6.5.  Accordingly, the \citet{riggs24} sample has a larger fraction of dwarfs that assemble their mass at earlier times.  Of the \citet{riggs24} simulated dwarfs with Log M$_{\star}$/M$_{\odot}$$>$6.5, 43\% assembled at least half of their cumulative mass by 10 Gyr ago, compared to only 12\%  for our observed sample.  

The aforementioned minor differences in both methodology and the distribution of target galaxy properties may be partially responsible for the somewhat tighter correlation we find between $\gamma_{90}$ and $\tau_{90}$(global) compared to either set of simulations (see Fig.~\ref{violinfig}).  Indeed, \citet{riggs24} posit that mass is the primary driver of age gradients, both directly and indirectly.  In a direct sense, mass determines the ability of a galaxy to drive the potential well fluctuations that are responsible for present-day outside-in gradients.  However, mass also plays an indirect role in setting age gradients, since only dwarfs with Log M$_{\star}$/M$_{\odot}$$\gtrsim$6 host star formation out to a maximum galactocentric radius that increases with time, flattening age gradients to an extent that depends on the global SFH of the galaxy.    

Lastly, we have made two simplifying assumptions to allow for a homogeneous analysis methodology across our sample.  First, we assume that the density profiles of our targets are exponential to enable the use of a single sample-wide method for placement of our radial bins, while there is evidence that some dwarfs may not be well fit by a single exponential \citep[e.g.,][]{battinelli07,higgs21a}.  Second, we have assumed that radial age gradients are linear and azimuthally symmetric.  While the quality of available data does not reveal any clear cases of non-linear age gradients such as the \enquote{V-shaped} radial trends in $\tau_{50}$ and $\tau_{90}$ seen at higher galaxy masses in the LMC and M33, there is at least one example of a non-linear age gradient detected in a dwarf within the present-day mass range we examine \citep{ruizlara21}.  Similarly, since our sample includes dwarfs with incomplete azimuthal coverage out to our chosen cut in galactocentric radius (i.e., $\mathcal{A}$$<$1 in Table \ref{propstab}), we have ignored the possibility of azimuthally-dependent radial stellar age gradients biasing our results.  While we found no significant correlations between our areal coverage fraction $\mathcal{A}$ versus $\gamma_{90}$ or $\gamma_{50}$, there is evidence for position-angle-dependent radial stellar age gradients in the mass range we examine.  Beyond the heavily interacting LMC and SMC, position-angle-dependent radial age gradients could be caused by either nascent spiral structure as suggested for NGC 6822 \citep{fusco14} and/or ram pressure stripping that could occur in environments that are more isolated than previously thought \citep{cohen25}.  Future modeling work may address the impact of observational biases in recovered radial age gradients in more detail using end-to-end simulations, while our intent here is to present and apply a stellar age gradient measurement technique that can be extended to additional targets regardless of morphology or incomplete spatial sampling.

\section{Conclusions and Future Work \label{futuresect}}

We have performed PSF photometry (Sect.~\ref{photsect}) of archival Hubble imaging of 42 Local Volume dwarf galaxies spanning a range of global properties including stellar mass, metallicity, morphology, environment and global SFH (Table \ref{propstab}).   
We calculated radial age gradients by spatially dividing each galaxy into radial bins extending out to 4.4$h_{r}$ (Sect.~\ref{structuralsect}), fitting SFHs independently to CMDs of each spatially selected subsample (Sect.~\ref{sfhsect}) and measuring $\gamma_{90}$ and $\gamma_{50}$, the gradients in $\tau_{50}$ and $\tau_{90}$ (the lookback times to form 50\% and 90\% of the cumulative stellar mass respectively), normalized to R$_{hl}$ (Fig.~\ref{sfh_example_fig}).  We compared the resulting database of observed $\gamma_{90}$ and $\gamma_{50}$ to predictions from multiple recent cosmological hydrodynamical simulations for the first time, and our principal findings are:
\begin{enumerate}

\item A search for correlations between $\gamma_{90}$ and the global galaxy properties we examine revealed a highly significant ($p$-values$\lesssim$0.001; Table \ref{coefftab90}) correlation between $\gamma_{90}$ and $\tau_{90}$(global) (see Fig.~\ref{obstaufig}), where $\tau_{90}$(global) is the galaxy-wide value of $\tau_{90}$ calculated by evaluating a linear fit to the radial gradient of per-ellipse $\tau_{90}$ values at R$_{hl}$ (see Sect.~\ref{calcsect}).  This correlation is consistent with predictions of two independent sets of recent cosmological hydrodynamical simulations (\citealt{graus19,riggs24}; Fig.~\ref{compsim90fig}) based on the interplay between feedback-driven stellar outward radial migration and an increasing maximum formation radius of young stars with time (Sect.~\ref{introsimsect}). 

\item In contrast to $\gamma_{90}$, $\gamma_{50}$ is not correlated with \textit{any} global galaxy properties we examined, including $\tau_{50}$(global) (Fig.~\ref{compsim50fig}, Table \ref{coefftab50}).  We demonstrated (Sect.~\ref{t50simsect}, Fig.~\ref{violinfig}) that the lack of a correlation between $\gamma_{50}$ and $\tau_{50}$(global) provides an independent method of observationally discriminating between the predictions of multiple sets of cosmological simulations, strongly supporting the predictions of \citet{riggs24} over those of \citet{graus19}.  Our results therefore argue against a scenario where feedback-induced potential fluctuations are accompanied by radially outflowing gas that imparts an initial radial velocity to young stars and sets a relationship between age gradients and global SFH prior to $\tau_{50}$(global).  

\item None of the dwarfs in our sample show statistically significant evidence for inside-out radial age gradients at the present day, in accord with the predictions of both sets of simulations examined.

\item We detect no correlation between age gradient slope and any of the environmental metrics examined ($\theta_{1}$, D(NN), D(NLG)$_{\rm 10}$; Tables \ref{coefftab90}-\ref{coefftab50}), implying that age gradients in dwarfs are driven mostly or entirely by internal processes, consistent with the \citet{riggs24} simulations.  

\end{enumerate}

We have demonstrated that in the presence of a sufficiently diverse and populous sample of target galaxies and a self-consistent analysis strategy, internal stellar age gradients in dwarfs provide a new avenue for observationally discriminating between the predictions of different sets of cosmological simulations.  However, our sample is neither unbiased nor complete (i.e., volume-limited), although these concerns may be addressed using a combination of existing imaging and the latest observational facilities.  For example, the \citet{riggs24} simulations predict that the lowest-mass dwarfs (Log M$_{\star}$/M$_{\odot}$$\lesssim$6) quenched shortly after reionization tend to have flat age gradients.  This prediction may be testable given high-resolution imaging of sufficient depth to combat statistical uncertainties on per-radial-bin SFHs, which are driven by the size of the stellar sample available in each radial bin.  

Beyond simply measuring age gradient slopes, the latest observational facilities may also bring within reach the possibility of characterizing non-linear radial age trends in dwarfs at sub-LMC masses.  There already exist individual cases demonstrating that high-fidelity observations can reveal both non-linearities and position-angle-dependent trends in spatially resolved SFHs of dwarfs \citep[e.g.,][also see the bottom row of Figs.~\ref{indslopes_parsec_t90fig} and \ref{indslopes_mist_t90fig}]{hidalgo13,ruizlara21,cohen25}.  Looking forward, the combination of spatial resolution, throughput and spatial coverage provided by next-generation imagers such as JWST, Roman and Euclid may provide yet more detailed constraints to discriminate between simulations with different input assumptions as we have done here.    

\begin{acknowledgements}

It is a pleasure to thank the anonymous referee for their thoughtful, unbiased review.  The data presented in this article were obtained from the Mikulski Archive for Space Telescopes (MAST) at the Space Telescope Science Institute. The specific observations analyzed can be accessed via \dataset[doi:10.17909/6fby-wa58]{https://doi.org/10.17909/6fby-wa58}.  
Support for this work was provided by NASA through grant AR-17038 from the Space Telescope Science Institute, which is operated by AURA, Inc., under NASA contract 5-26555.  The data presented in this article were obtained from the Mikulski Archive for Space Telescopes (MAST) at the Space Telescope Science Institute.  This research has made use of the NASA Astrophysics Data System Bibliographic Services.  R.~E.~C. acknowledges additional support from Rutgers the State University of New Jersey.  O.~G.~T. acknowledges support from grant AR-16155 from the Space Telescope Science Institute and from a Carnegie-Princeton Fellowship through Princeton University and the Carnegie Observatories.

\end{acknowledgements}

\facilities{HST - Hubble Space Telescope satellite (ACS/WFC, WFC3, WFPC2)}

\software{astropy \citep{astropy}, matplotlib \citep{matplotlib}, numpy \citep{numpy}, emcee \citep{emcee}, DOLPHOT \citep{dolphot,dolphot2}, MATCH \citep{match}}

\bibliography{RadGrad}
\bibliographystyle{aasjournalv7}

\appendix

\section{Testing the Impact of Different Assumptions on Radial Gradients \label{testsect}}

\begin{figure}
\gridline{\fig{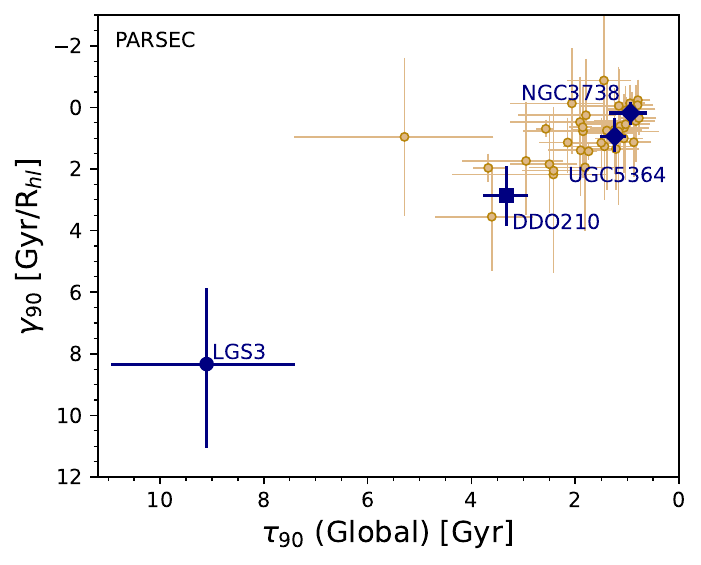}{0.45\textwidth}{}
          \fig{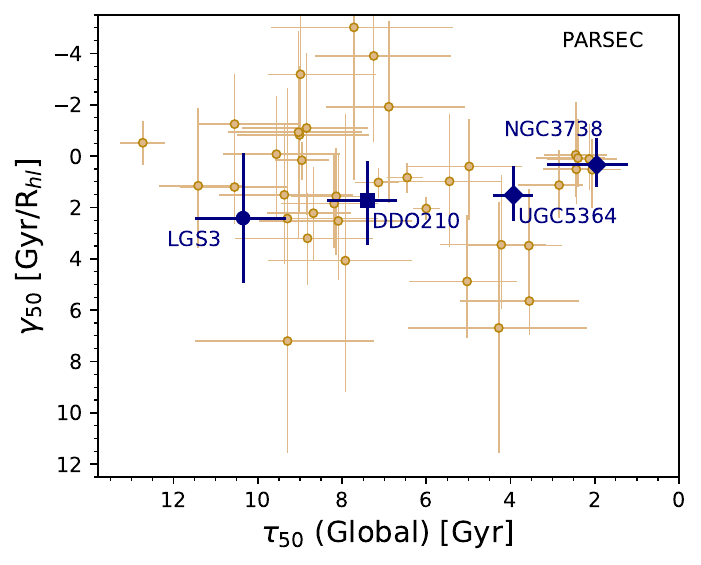}{0.45\textwidth}{}}
\gridline{\fig{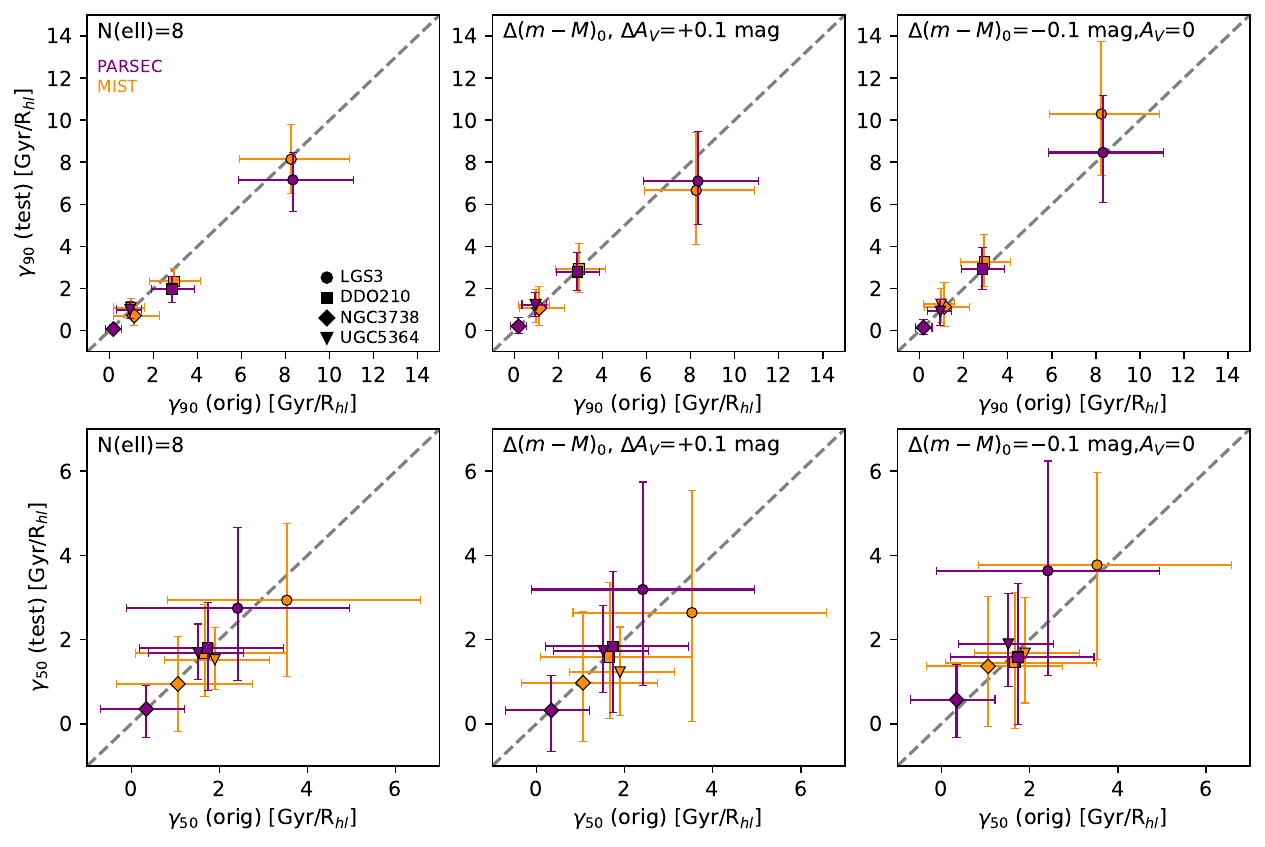}{0.9\textwidth}{}}
\caption{\textbf{Top row:} As in Fig.~\ref{obstaufig}, but illustrating in blue the subset of target galaxies used to test the sensitivity of our age gradient slopes to different input assumptions (see text for details).  \textbf{Middle Row:} Comparison of age gradient slopes $\gamma_{90}$ resulting from changes to our input assumptions shown on the vertical axis against the values reported in Table \ref{slopetab} on the horizontal axis.  The dashed grey line indicates equality.  Each panel represents a different change to our input assumptions for measuring radial gradient slopes, labeled at the top of each panel.  Points are color-coded to indicate the stellar evolutionary models used for SFH fitting and different symbols correspond to the different galaxies selected as test cases highlighted in the top row.  \textbf{Bottom Row}: Same as the middle row, but comparing $\gamma_{50}$ rather than $\gamma_{90}$.  In all cases, the measured age gradient slopes are unaffected beyond their uncertainties.
\label{testcasefig}}
\end{figure}

We select a subset of our targets to test the sensitivity of our age gradient slopes $\gamma_{90}$ and $\gamma_{50}$ to several different input assumptions made in our analysis in Sect.~\ref{methodsect}.  The galaxies selected as test cases, overplotted on our full sample in the top row of Fig.~\ref{testcasefig}, were chosen based on two factors: First, they have small relative uncertainties on their age gradient slopes compared to other galaxies in the sample, allowing the most stringent tests of the sensitivity of our age gradients to the different assumptions described below.  Second, they were chosen to span the range of observed age gradient slopes in case the extent to which our assumptions affect our slope measurements is related to the age gradient slopes themselves.  

For each galaxy, we altered three different input assumptions before remeasuring age gradients, which we compare to our reported values (listed in Table \ref{slopetab}) in the middle (for $\gamma_{90}$) and bottom (for $\gamma_{50}$) rows of Fig.~\ref{testcasefig}.  First, we double the number of radial bins from four to eight (left column in the lower two rows of Fig.~\ref{testcasefig}).  The next two tests consist of increasing (middle column) and decreasing (right-hand column) the assumed distance modulus $(m-M)_{0}$ and foreground extinction A$_{V,fg}$ by $\pm$0.1 mag\footnote{Since the vast majority of our sample, including all of the selected test case galaxies, have A$_{V,fg}$$\leq$0.1 mag (see Table \ref{propstab}), in practice we set A$_{V,fg}$=0 for the test cases with $\Delta$A$_{V,fg}$$=$$-$0.1 to avoid negative values of A$_{V,fg}$.} to test the sensitivity of our age gradients to assumptions on these values.  
As illustrated in the middle and bottom rows of Fig.~\ref{testcasefig}, we found that none of these changes affected our age gradient slopes beyond their uncertainties, and we also found (although not shown here for clarity) that varying either $(m-M)_{0}$ or A$_{V,fg}$ separately in turn by $\pm$0.1 mag also did not affect age gradients beyond their uncertainties.

\section{Correlation Coefficients: Age Gradient Slopes Versus Global Parameters \label{coefftabsect}}

We present the Pearson linear correlation coefficients and Spearman rank correlation coefficients between the radial age gradient slope $\gamma_{90}$ and $\gamma_{50}$ versus seven global properties of our target galaxies (described in Sect.~\ref{resultsect}) below in Tables \ref{coefftab90} and \ref{coefftab50} respectively. 

\begin{deluxetable}{lcccc}[h]
\tabletypesize{\small} 
\tablecaption{Correlation Coefficients: $\gamma_{90}$ Versus Global Galaxy Properties \label{coefftab90}}
\tablehead{\colhead{Parameter} & \colhead{Pearson $\rho$} & \colhead{Log$_{\rm 10}$ $p$-value} & \colhead{Spearman $r_{s}$} & \colhead{Log$_{\rm 10}$ $p$-value}}
\startdata
\hline
\multicolumn{5}{c}{Assuming PARSEC Models} \\
\hline
$\tau_{90}$(Global)      & 0.56$_{-0.13}^{+0.12}$ & -3.90$_{-2.28}^{+1.59}$ & 0.35$_{-0.13}^{+0.13}$ & -1.64$_{-1.21}^{+0.83}$ \\
12+Log(O/H)  & -0.07$_{-0.14}^{+0.14}$ & -0.26$_{-0.42}^{+0.20}$ & -0.09$_{-0.14}^{+0.15}$ & -0.32$_{-0.50}^{+0.24}$ \\
Log M$_{\star}$/M$_{\odot}$  & -0.43$_{-0.08}^{+0.08}$ & -2.39$_{-0.93}^{+0.75}$ & -0.38$_{-0.10}^{+0.11}$ & -1.90$_{-1.04}^{+0.79}$ \\
M$_{\rm HI}$/M$_{\star}$ & -0.10$_{-0.10}^{+0.11}$ & -0.30$_{-0.42}^{+0.23}$ & -0.18$_{-0.11}^{+0.13}$ & -0.62$_{-0.64}^{+0.44}$ \\
$\theta_{1}$     & 0.25$_{-0.11}^{+0.11}$ & -0.96$_{-0.76}^{+0.54}$ & 0.18$_{-0.12}^{+0.12}$ & -0.58$_{-0.67}^{+0.41}$ \\
D(NN)     & -0.09$_{-0.09}^{+0.09}$ & -0.27$_{-0.34}^{+0.20}$ & -0.08$_{-0.12}^{+0.12}$ & -0.29$_{-0.43}^{+0.22}$ \\
D(NLG$_{\rm 10}$)  & -0.23$_{-0.08}^{+0.09}$ & -0.85$_{-0.48}^{+0.41}$ & -0.17$_{-0.11}^{+0.11}$ & -0.57$_{-0.60}^{+0.39}$ \\
\hline
\multicolumn{5}{c}{Assuming MIST Models} \\
\hline
$\tau_{90}$(Global)     & 0.45$_{-0.15}^{+0.13}$ & -2.53$_{-1.67}^{+1.28}$ & 0.26$_{-0.14}^{+0.12}$ & -1.03$_{-0.89}^{+0.67}$ \\
12+Log(O/H)  & 0.05$_{-0.15}^{+0.15}$ & -0.27$_{-0.40}^{+0.20}$ & 0.05$_{-0.16}^{+0.16}$ & -0.29$_{-0.45}^{+0.23}$ \\
Log M$_{\star}$/M$_{\odot}$   & -0.30$_{-0.09}^{+0.11}$ & -1.30$_{-0.75}^{+0.62}$ & -0.24$_{-0.12}^{+0.12}$ & -0.87$_{-0.79}^{+0.52}$ \\
M$_{\rm HI}$/M$_{\star}$ & -0.21$_{-0.10}^{+0.11}$ & -0.73$_{-0.58}^{+0.46}$ & -0.30$_{-0.13}^{+0.14}$ & -1.30$_{-1.08}^{+0.76}$ \\
$\theta_{1}$   & 0.21$_{-0.12}^{+0.11}$ & -0.74$_{-0.68}^{+0.47}$ & 0.13$_{-0.12}^{+0.12}$ & -0.43$_{-0.55}^{+0.32}$ \\
D(NN)     & -0.09$_{-0.11}^{+0.12}$ & -0.30$_{-0.42}^{+0.23}$ & -0.12$_{-0.13}^{+0.13}$ & -0.37$_{-0.61}^{+0.28}$ \\
D(NLG$_{\rm 10}$) & -0.03$_{-0.10}^{+0.11}$ & -0.20$_{-0.32}^{+0.16}$ & 0.02$_{-0.13}^{+0.12}$ & -0.23$_{-0.37}^{+0.16}$ \\
\enddata
\end{deluxetable}

\begin{deluxetable}{lcccc}[h]
\tabletypesize{\small} 
\tablecaption{Correlation Coefficients: $\gamma_{50}$ Versus Global Galaxy Properties \label{coefftab50}}
\tablehead{\colhead{Parameter} & \colhead{Pearson $\rho$} & \colhead{Log$_{\rm 10}$ $p$-value} & \colhead{Spearman $r_{s}$} & \colhead{Log$_{\rm 10}$ $p$-value}}
\startdata
\hline
\multicolumn{5}{c}{Assuming PARSEC Models} \\
\hline
$\tau_{50}$(Global)    & -0.08$_{-0.12}^{+0.13}$ & -0.29$_{-0.40}^{+0.21}$ & -0.07$_{-0.12}^{+0.13}$ & -0.30$_{-0.42}^{+0.23}$ \\
12+Log(O/H)  & 0.14$_{-0.13}^{+0.11}$ & -0.39$_{-0.41}^{+0.29}$ & 0.11$_{-0.13}^{+0.12}$ & -0.30$_{-0.44}^{+0.22}$ \\
Log M$_{\star}$/M$_{\odot}$ & -0.02$_{-0.09}^{+0.08}$ & -0.15$_{-0.25}^{+0.11}$ & -0.03$_{-0.11}^{+0.11}$ & -0.21$_{-0.30}^{+0.16}$ \\
M$_{\rm HI}$/M$_{\star}$ & -0.16$_{-0.12}^{+0.13}$ & -0.50$_{-0.62}^{+0.37}$ & -0.13$_{-0.12}^{+0.12}$ & -0.39$_{-0.58}^{+0.29}$ \\
$\theta_{1}$     & 0.01$_{-0.11}^{+0.11}$ & -0.21$_{-0.31}^{+0.16}$ & -0.03$_{-0.11}^{+0.12}$ & -0.23$_{-0.29}^{+0.16}$ \\
D(NN)     & 0.05$_{-0.10}^{+0.09}$ & -0.18$_{-0.28}^{+0.13}$ & 0.02$_{-0.11}^{+0.11}$ & -0.20$_{-0.32}^{+0.15}$ \\
D(NLG$_{\rm 10}$)  & 0.03$_{-0.09}^{+0.09}$ & -0.15$_{-0.23}^{+0.11}$ & 0.05$_{-0.11}^{+0.12}$ & -0.23$_{-0.36}^{+0.16}$ \\
\hline
\multicolumn{5}{c}{Assuming MIST Models} \\
\hline
$\tau_{50}$(Global)   & 0.13$_{-0.13}^{+0.14}$ & -0.41$_{-0.67}^{+0.30}$ & 0.10$_{-0.14}^{+0.13}$ & -0.33$_{-0.52}^{+0.25}$ \\
12+Log(O/H)  & -0.04$_{-0.14}^{+0.13}$ & -0.24$_{-0.38}^{+0.18}$ & -0.04$_{-0.14}^{+0.14}$ & -0.24$_{-0.40}^{+0.19}$ \\
Log M$_{\star}$/M$_{\odot}$ & -0.12$_{-0.09}^{+0.10}$ & -0.37$_{-0.39}^{+0.27}$ & -0.13$_{-0.11}^{+0.12}$ & -0.42$_{-0.52}^{+0.29}$ \\
M$_{\rm HI}$/M$_{\star}$ & -0.04$_{-0.12}^{+0.12}$ & -0.23$_{-0.33}^{+0.17}$ & -0.08$_{-0.12}^{+0.13}$ & -0.28$_{-0.43}^{+0.21}$ \\
$\theta_{1}$     & 0.08$_{-0.12}^{+0.12}$ & -0.27$_{-0.44}^{+0.19}$ & 0.04$_{-0.12}^{+0.12}$ & -0.23$_{-0.35}^{+0.17}$ \\
D(NN)     & -0.01$_{-0.10}^{+0.11}$ & -0.19$_{-0.30}^{+0.14}$ & -0.00$_{-0.12}^{+0.11}$ & -0.21$_{-0.33}^{+0.16}$ \\
D(NLG$_{\rm 10}$) & 0.08$_{-0.11}^{+0.10}$ & -0.27$_{-0.38}^{+0.20}$ & 0.11$_{-0.12}^{+0.13}$ & -0.34$_{-0.55}^{+0.26}$ \\
\enddata
\end{deluxetable}

\section{Radial Age Gradients: Linear Fits \label{indslopesect}}

\begin{figure}
\gridline{\fig{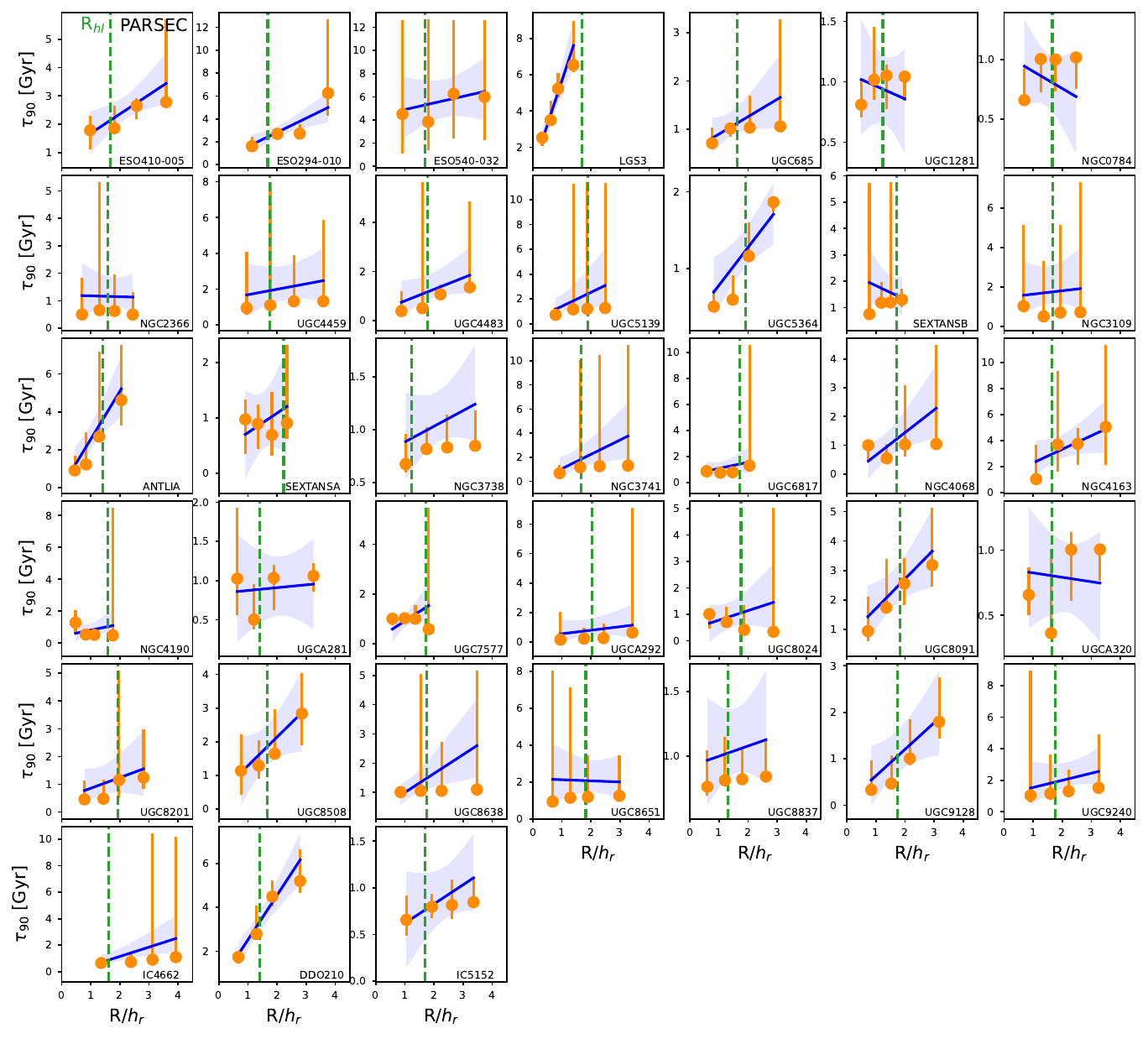}{0.99\textwidth}{}}
\vspace{-1.1cm}
\gridline{\fig{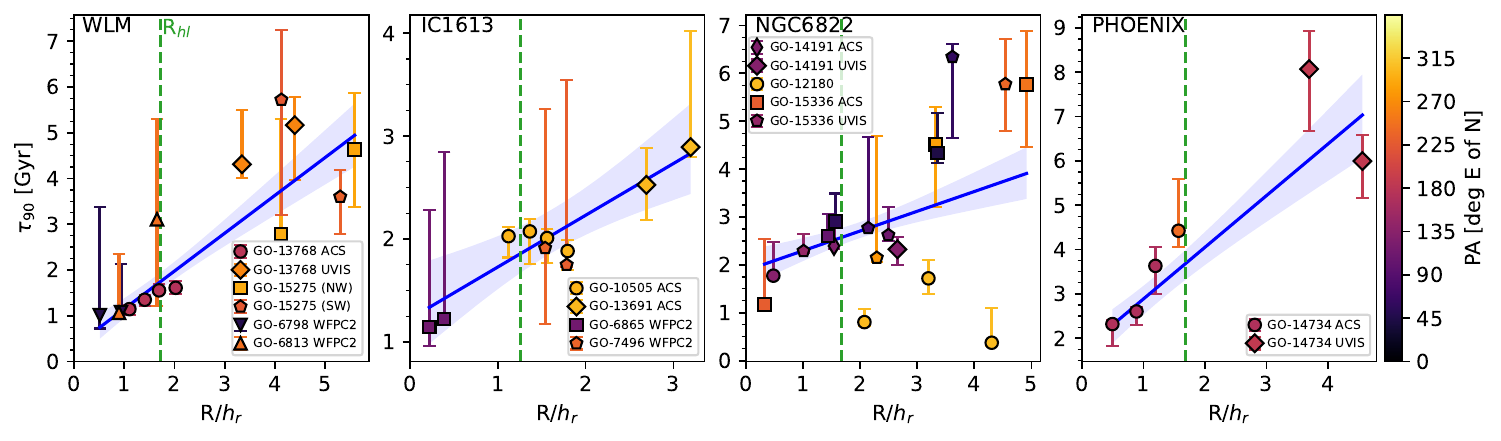}{0.99\textwidth}{}}
\vspace{-1cm}
\caption{Best-fit radial stellar age gradient slopes of $\tau_{90}$ versus R$_{\rm SMA}$ shown for each individual target assuming PARSEC evolutionary models.  In the bottom row, we show linear fits for the four targets with multiple non-overlapping pointings available in the archive, where different symbols indicate different observing programs and color-coding indicates position angle relative to the galaxy center corresponding to the colorbar on the bottom right. 
\label{indslopes_parsec_t90fig}}
\end{figure}

\begin{figure}
\gridline{\fig{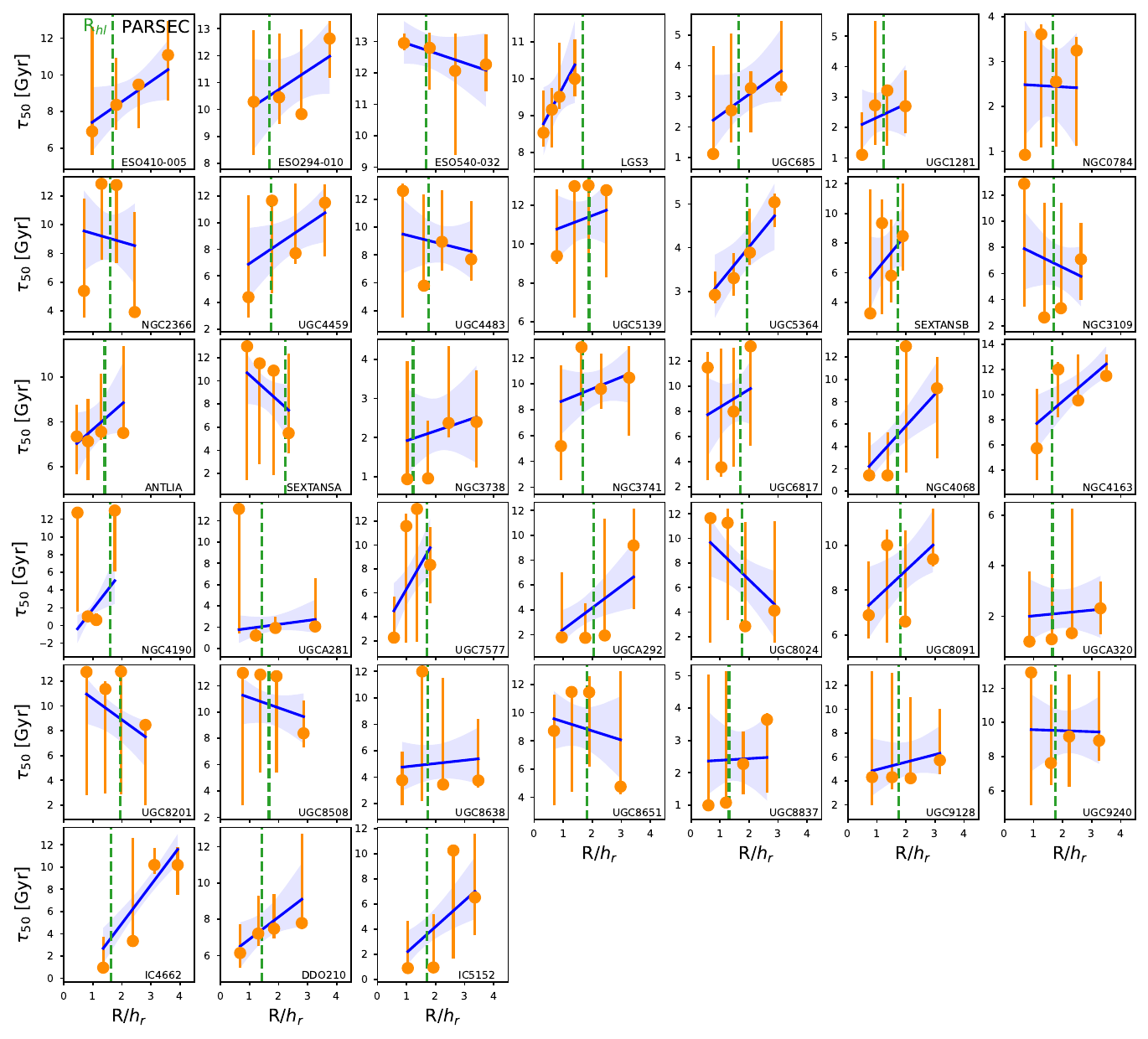}{0.99\textwidth}{}}
\vspace{-1cm}
\gridline{\fig{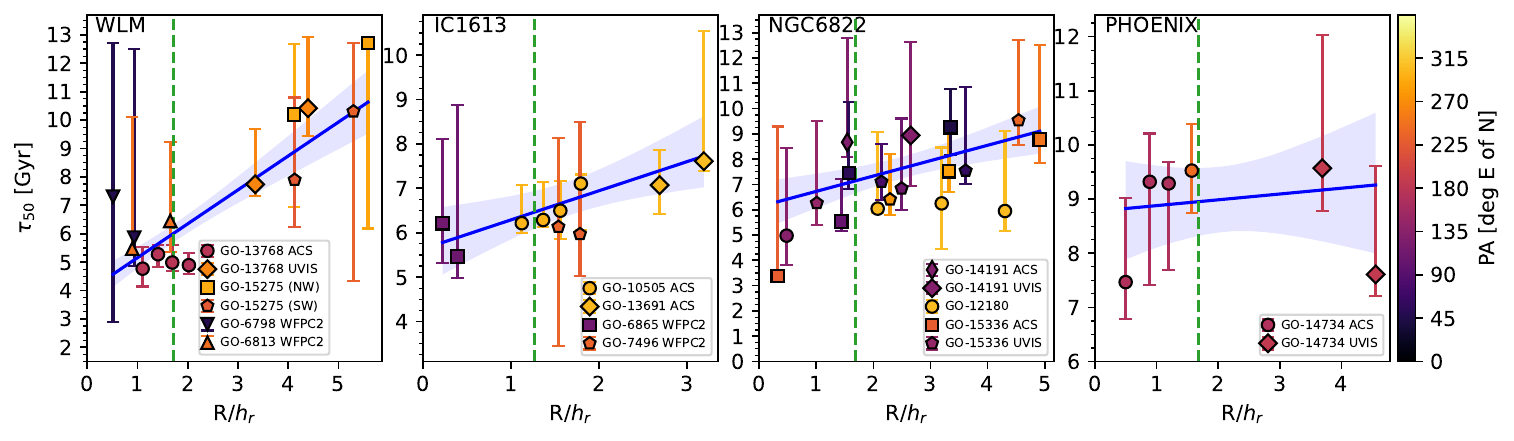}{0.99\textwidth}{}}
\vspace{-1cm}
\caption{Best-fit radial stellar age gradient slopes of $\tau_{50}$ versus R$_{\rm SMA}$ shown for each individual target assuming PARSEC evolutionary models.  
\label{indslopes_parsec_t50fig}}
\end{figure}

\begin{figure}
\gridline{\fig{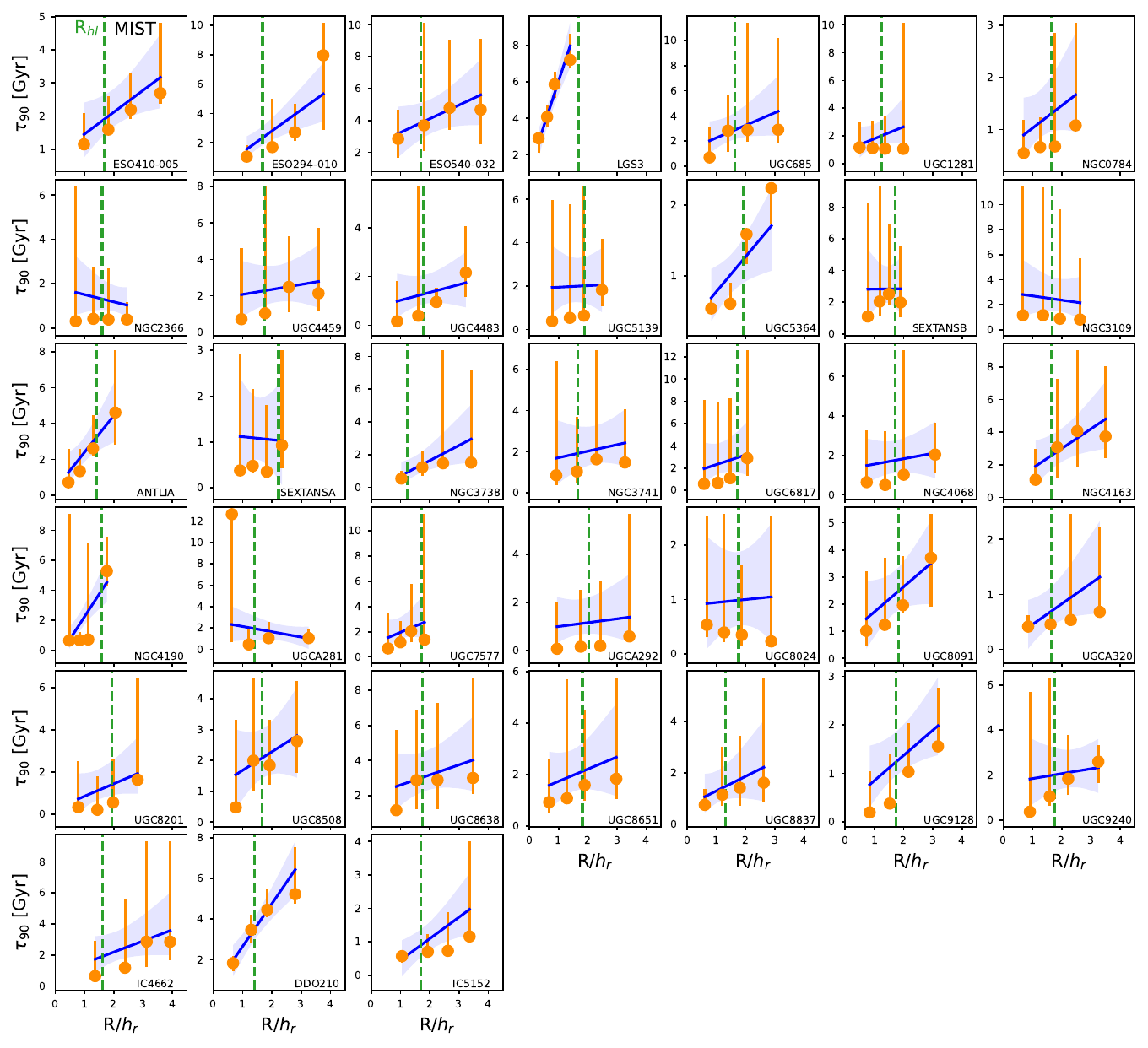}{0.99\textwidth}{}}
\vspace{-1cm}
\gridline{\fig{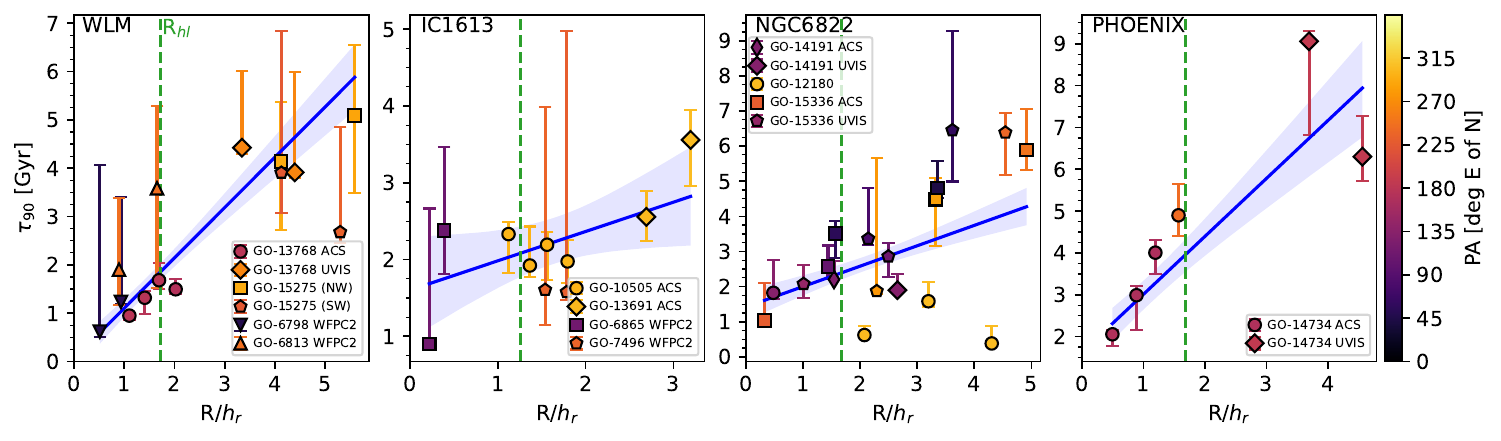}{0.99\textwidth}{}}
\vspace{-1cm}
\caption{As in Fig.~\ref{indslopes_parsec_t90fig} but assuming MIST evolutionary models.  
\label{indslopes_mist_t90fig}}
\end{figure}

\begin{figure}
\gridline{\fig{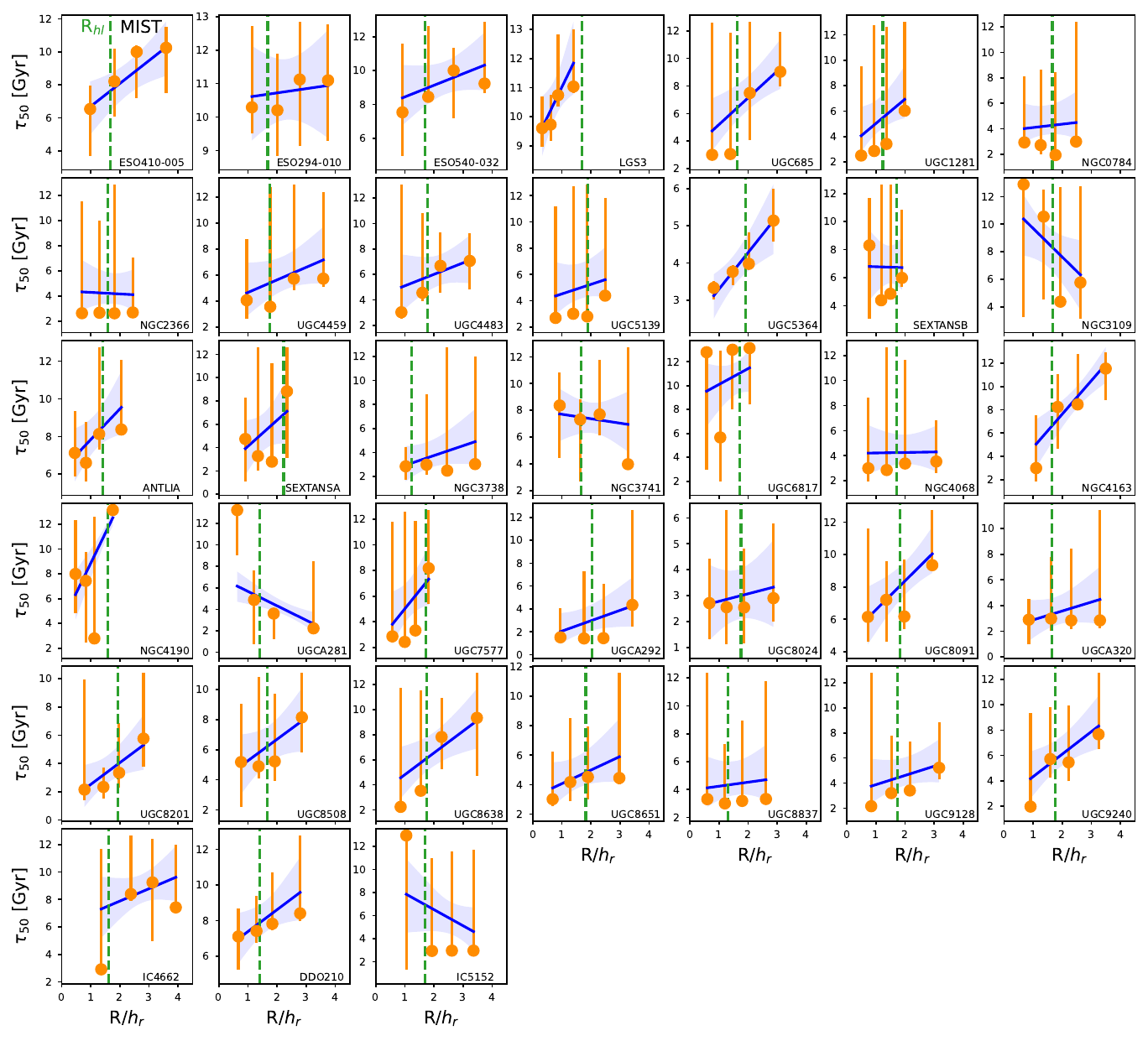}{0.99\textwidth}{}}
\vspace{-1cm}
\gridline{\fig{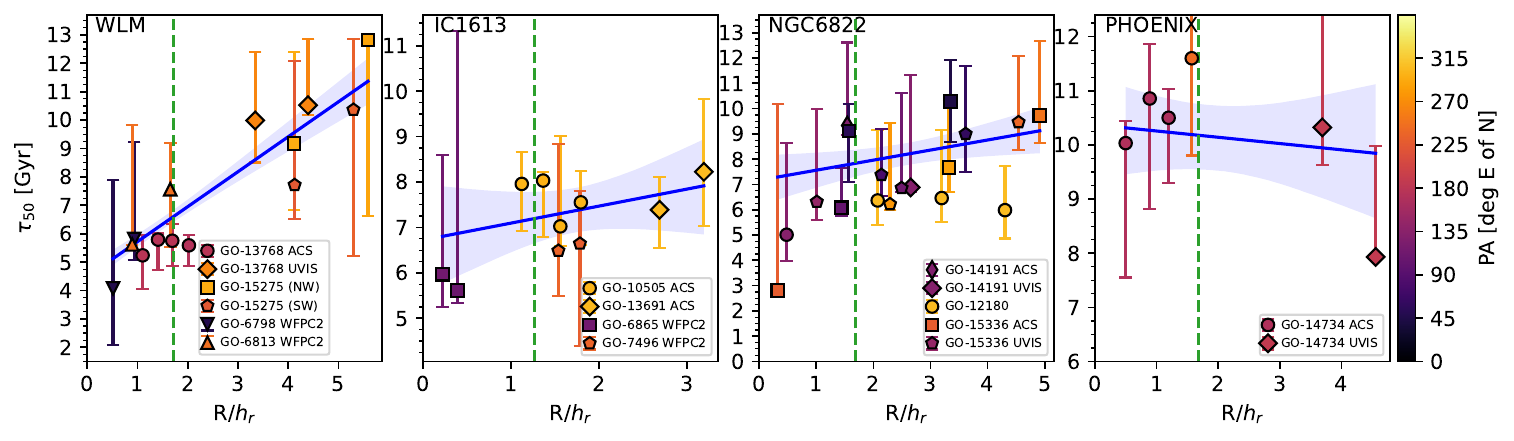}{0.99\textwidth}{}}
\vspace{-1cm}
\caption{As in Fig.~\ref{indslopes_parsec_t50fig} but assuming MIST evolutionary models. 
\label{indslopes_mist_t50fig}}
\end{figure}

The best-fit radial gradient slopes $\gamma_{90}$ and $\gamma_{50}$ (and 1-$\sigma$ uncertainties indicated by shading) are shown for all of our target galaxies individually in Figs.~\ref{indslopes_parsec_t90fig}-\ref{indslopes_parsec_t50fig} respectively assuming PARSEC stellar evolutionary models for SFH fitting, and Figs.~\ref{indslopes_mist_t90fig}-\ref{indslopes_mist_t50fig} respectively assuming MIST stellar evolutionary models.  In the bottom row of each figure, we show our linear age gradient fits for the four galaxies with multiple non-overlapping pointings available (see Sect.~\ref{speccasesect}), where different symbols correspond to the various observing programs available while color-coding indicates the position angle of each radial bin of each pointing relative to the target center location (see Table \ref{slopetab}). 

\end{document}